\documentclass[smallextended,runningheads]{svjour3}
\usepackage{graphicx}

\usepackage{graphicx} 
\usepackage{amsmath}
\usepackage{amsfonts}
\usepackage{latexsym}
\usepackage{amssymb}

\newcommand{\wt}{\widetilde}
\newcommand{\calf}{{\cal F}}

\newcommand{\caln}{{\cal N}}
\newcommand{\caly}{{\cal Y}}
\newcommand{\del}{{\delta}}
\newcommand{\ep}{\epsilon}
\newcommand{\nn}{\nonumber}
\newcommand{\be}{\begin{equation}}
\newcommand{\ee}{\end{equation}}
\newcommand{\bea}{\begin{eqnarray}}
\newcommand{\eea}{\end{eqnarray}}
\newcommand{\beas}{\begin{eqnarray*}}
\newcommand{\eeas}{\end{eqnarray*}}
\newcommand{\mbb}{\mathbb}

\newcommand{\part}{\partial}

\begin{document}

\title{Momentum-Space Approach to Asymptotic Expansion  for Stochastic Filtering~\footnote{The manuscript is accepted for publication in {\it the Annals of the Institute of Statistical Mathematics}. 
The official manuscript will be available in http://www.ism.ac.jp/editsec/aism-e.html.}
}

\titlerunning{Asymptotic Expansion for Stochastic Filtering}        

\author{Masaaki Fujii
}


\institute{Masaaki Fujii \at
              Faculty of Economics, The University of Tokyo, 7-3-1, Hongo,  Bunkyo-ku, Tokyo, 113-0033 Japan\\
              \email{mfujii@e.u-tokyo.ac.jp}                       
}

\date{1st version: September 10,  2012/ Current version: March 24, 2013}

\maketitle
\begin{abstract}
This paper develops an asymptotic expansion technique in momentum space for stochastic filtering.  
It is shown that Fourier transformation combined with a polynomial-function approximation of
the  nonlinear terms gives a closed recursive system of ordinary differential equations (ODEs)
for the relevant conditional distribution.
Thanks to the simplicity of the ODE system, higher order calculation can be performed easily.
Furthermore, solving ODEs sequentially with small sub-periods with updated 
initial conditions makes it possible to implement a {\it substepping method} for asymptotic expansion in
a numerically efficient way.
This is found to improve the performance significantly where otherwise the approximation fails badly.
The method is expected to provide a useful tool for more realistic financial modeling with unobserved parameters,
and also for problems involving nonlinear measure-valued processes.

\end{abstract}

\section{Introduction}
\label{intro}
In many areas, researchers frequently encounter the situation where crucial parameters 
for their models are not directly observable in our mother nature or in experiments. This is particularly the case, for example,
in engineering, applied physics, finance and economics.
To get the best estimate of the unobservable from what we can directly observe is the goal of stochastic filtering.
The most famous example with analytical solution is Kalman-Bucy filter [Bucy (1959), Kalman (1960), Kalman and Bucy (1961)], which assumes that both of the signal and observation processes are linear and hence is associated with 
Gaussian distribution.

However, there are many cases where interested variables  follow nonlinear 
stochastic processes and their distributions are far from Gaussian.
This is particularly the case for financial problems. In fact, many 
people were forced to realize the sheer impacts of non-Gaussianity in the last financial crisis
followed by the collapse of Lehman Brothers.
Researchers, practitioners as well as regulators now clearly recognize the importance 
of not only the first two moments but also every other detail of the relevant distribution.
Here, we need to deal with nonlinear filtering problems.
Filtering theory has a long history and is still developing very rapidly,
partly helped by the great increase of computational power.
Recently, there appeared a thick volume edited by Crisan and Rozovski$\check{\mbox{i}}$ (2011) from Oxford university press,
which contains latest developments and reviews for theoretical as well as numerical techniques for 
nonlinear filtering problems. In particular, the article written by Frey and Runggaldier in the same 
volume gives a nice review of applications to various financial valuation problems under
partial observation.


In this paper, we propose a simple approximation scheme based on an asymptotic expansion method
in momentum space for nonlinear filtering problems. The method should be 
also useful for other financial problems that do not require filtering.
Widely used "position-space" asymptotic expansion method~(See, Takahashi (1999), and references therein.)
is transformed into a simpler form in the momentum (or Fourier transformed) apace, 
and the resultant dynamics of the characteristic function is given by 
a closed recursive system of ordinary differential equations (ODEs).
It is shown that the form of ODEs unchanged for any order of expansion, which allows 
straightforward numerical implementation for higher order approximations.
Furthermore, dividing the original time horizon into a set of small sub-periods 
and solving the ODEs sequentially with updated initial conditions, which we call {\it substepping method} for 
asymptotic expansion, increase the parameter space where the approximation is effective.
Two simple examples are discussed to demonstrate how the method works.
We also make a brief comment on the possibility that the same method can be used to analyze
other measure-valued nonlinear systems.

\section{Preliminaries for Nonlinear Filtering}
\subsection{Zakai equation}
\label{sec-zakai}
Let $(\Omega,\calf,\mbb{P})$ be a probability space with a filtration $(\calf_t)_{t\geq 0}$
satisfying the usual conditions.
We consider  $n$-dimensional signal process $X=\{X_t,t\geq 0\}$ and
$m$-dimensional observation process $Y=\{Y_t,t\geq 0\}$ following the dynamics of
\bea
\label{signal_org}
&&dX_t=\mu_t(X_t)dt+\eta_t(X_t)dV_t+\bar{\eta}_t(X_t)dW_t\\
&&dY_t=h_t(X_t)dt+dW_t
\label{obs_org}
\eea
with $Y_0=0$ and an independent initial distribution for $X_0$. 
Here $V$ and $W$ are independent $(\mbb{P},\calf)$-Brownian motions
with dimensionality $d$ and $m$, respectively. 
$\mu$, $h$, $\eta$ and $\bar{\eta}$ are deterministic functions of 
$(t,x)$ and take values
in $\mbb{R}^n, \mbb{R}^m, \mbb{R}^{n\times d}$ and $\mbb{R}^{n\times m}$, respectively.
The functions $\mu$, $\eta$ and $\bar{\eta}$ are assumed to satisfy appropriate conditions
so that (\ref{signal_org}) has a unique solution.
The measurable function $h$ is assumed to satisfy the conditions 
that make the following process $Z=\{Z_t, t\geq 0\}$ be a martingale:
\bea
Z_t=\exp\left(-\int_0^t h_s(X_s)^\top dW_s-\frac{1}{2}\int_0^t ||h_s(X_s)||^2 ds\right) \nn
\eea
where $\top$ denotes the transposition.
We denote $\{\caly_t, t\geq 0\}$ be the usual augmented filtration generated by the process $Y$.
Our goal of the filtering problem is to obtain the conditional distribution $\pi_t$ of the signal $X$
at time $t$ given the information available from observing the process $Y$ in the interval of $[0,t]$.
In other words, for a given arbitrary bounded function $\varphi$, compute
\bea
\pi_t (\varphi)=\mathbb{E}\Bigl[\varphi(X_t)\Bigr|\caly_t\Bigr]~. \nn
\eea

Let us define the measure $\wt{\mbb{P}}$ by
\bea
\left.\frac{d\wt{\mbb{P}}}{d\mbb{P}}\right|_{\calf_t}=Z_t \nn 
\eea
and the associated inverse relation
\bea
\left.\frac{d\mbb{P}}{d\wt{\mbb{P}}}\right|_{\calf_t}=\wt{Z}_t \nn
\eea
where $\wt{Z}_t=Z_t^{-1}$ can be written as
\bea
\wt{Z}_t=\exp\left(\int_0^t h_s(X_s)^\top dY_s-\frac{1}{2}\int_0^t ||h_s(X_s)||^2 ds\right)~. \nn
\eea
Note that the process $Y$ becomes a standard $(\wt{\mbb{P}},\calf)$-Brownian motion.

We define the unnormalized conditional distribution of $X$ to be 
the measure-valued process $\rho=\{\rho_t, t\geq 0\}$
\bea
\rho_t(\varphi)=\wt{\mbb{E}}\Bigl[\wt{Z}_t\varphi(X_t)\Bigr|\caly_t\Bigr]\qquad  \wt{\mbb{P}}-a.s. \nn
\eea
which is $\caly_t$-adapted and c\`adl\`ag. Here, $\wt{\mbb{E}}[\cdot]$ denotes the 
expectation in the measure $\wt{\mbb{P}}$.
The desired filtered density function can then be obtained from the relation
\bea
\pi_t(\varphi)=\frac{\rho_t(\varphi)}{\rho_t(\bf{1})}. \nn
\eea
It is known that the dynamics of $\rho$ satisfies the following {\it Zakai} equation
for arbitrary smooth bounded function $\varphi$:
\bea
\rho_t(\varphi)=\rho_0(\varphi)+\int_0^t \rho_s(A_s\varphi)ds+\int_0^t \rho_s\Bigl((h_s+B_s)^\top \varphi\Bigr)dY_s
\label{zakai}
\eea
with initial value $\rho_0(\varphi)=\mbb{E}[\varphi(X_0)]$ associated with a
given distribution of $X_0$.
Here, $A_s$ is the infinitesimal generator of $X$ at time $s$ 
\bea
A_s=\sum_{i=1}^n \mu^i_s(x)\frac{\part}{\part x_i}+\frac{1}{2}\sum_{i,j=1}^n
\bigl(\eta_s\eta_s^\top(x)+\bar{\eta}_s\bar{\eta}_s^\top(x)\bigr)_{i,j}\frac{\part^2}{\part x_i\part x_j} \nn
\eea
and 
\bea
B_s^k=\sum_{i=1}^n(\bar{\eta}_s^\top(x))_{k,i}\frac{\part}{\part x_i}, \qquad k=1, \cdots, m. \nn
\eea
For the derivation of the Zakai equation and the other technical details, see Bain and Crisan (2008), for example.
The goal of this paper is to develop a simple scheme to solve the Zakai equation $(\ref{zakai})$.

\subsection{Filtered Characteristic Function}
\label{c-function}
Let us consider a function
\bea
\psi(\xi,x)=\exp\Bigl(i\xi^\top x\Bigr) \nn
\eea
with $\xi,x~\in\mbb{R}^n$, where $i=\sqrt{-1}$.
If one obtains the conditional expectation of this function, i.e., 
\bea
\pi_t(\psi(\xi,\cdot))=\frac{\wt{\mbb{E}}\Bigl[\exp\Bigl(i\xi^\top X_t\Bigr)\wt{Z}_t\Bigr|\caly_t\Bigr]}{
\wt{\mbb{E}}\bigl[\wt{Z}_t\bigr|\caly_t\bigr]} \nn
\eea
for each $\xi$, one can derive the conditional expectation for 
an arbitrary choice of $\varphi$. This fact can be seen as follows:
Let us consider the inverse Fourier transformation $\phi_t(\cdot)$
\bea
\phi_t(z)=\frac{1}{(2\pi)^n}\int_{\mbb{R}^n}e^{-i\xi^\top z}\pi_t(\psi(\xi,\cdot))d^n\xi \nn
\eea
which can be evaluated as
\bea
\phi_t(z)&=&\frac{1}{(2\pi)^n}\int_{\mbb{R}^d}\frac{1}{\wt{\mbb{E}}\bigl[\wt{Z}_t\bigr|\caly_t\bigr]}
\wt{\mbb{E}}\Bigl[\exp\Bigl(i\xi^\top(X_t-z)\Bigr)\wt{Z}_t\Bigr|\caly_t\Bigr]d^n\xi\nn \\
&=&\frac{\wt{\mbb{E}}\Bigl[\del(X_t-z)\wt{Z}_t\Bigr|\caly_t\Bigr]}{\wt{\mbb{E}}\bigl[\wt{Z}_t\bigr|\caly_t\bigr]}~, \nn
\eea
where $\del(\cdot)$ denotes a $n$-dimensional Dirac delta function.
The above function actually corresponds to the conditional density of the $X_t$ since
\bea
\int_{\mbb{R}^d}\varphi(z)\phi_t(z)d^n z&=&\frac{\wt{\mbb{E}}\Bigl[\varphi(X_t)\wt{Z}_t\Bigr|\caly_t\Bigr]}{
\wt{\mbb{E}}\bigl[\wt{Z}_t\bigr|\caly_t\bigr]} \nn \\
&=&\pi_t(\varphi)~. \nn
\eea
Therefore, $\{\pi_t(\psi(\xi))\}$ and equivalently $\{\rho_t(\psi(\xi))\}$ contain
all the important information one needs.
In fact, many of the financial valuation problems with partial information end up with 
calculating an integration with the conditional density $\phi_t(dz)$ (See, Frey and Runggaldier (2011) for a review.).

\section{Asymptotic Expansion in Momentum Space}
From the discussion in the previous section, we want to solve
the Zakai equation for the characteristic function $\rho_t(\psi(\xi))$.
Although one can directly approximate the Zakai equation for a specific problem,
it would be uneconomical. This is because the form of Zakai equation depends on the target function $\varphi$
for which one takes conditional expectation even in a common signal-observation system.

\subsection{Perturbed System}
\label{sec-perturbed}
In order to make the system tractable, we replace 
the parameters in the original system in (\ref{signal_org}) and (\ref{obs_org}) by
\bea
&&\mu_t(x)\rightarrow f_t+\ep F_t(x) \qquad \eta_t(x)\rightarrow \nu_t+\ep \sigma_t(x) \nn \\
&&\bar{\eta}_t(x)\rightarrow \ep \gamma_t(x) \qquad\qquad~ h_t(x)\rightarrow \ep H_t(x)\nn ~
\eea
and consider 
the $n$-dimensional signal and the $m$-dimensional observation processes as
\bea
&&dX_t^{(\ep)}=\Bigl(f_t+\ep F_t(X_t^{(\ep)})\Bigr)dt+\Bigl(\nu_t+\ep \sigma_t(X_t^{(\ep)})\Bigr)dV_t
+\ep \gamma_t(X_t^{(\ep)})dW_t  \nn \\
&& dY_t=\ep H_t(X_t^{(\ep)})dt+dW_t~. \nn 
\eea
Here, $f_t\in\mbb{R}^n$ and $\nu_t\in \mbb{R}^{n\times d}$ are deterministic functions of time and
$\nu_t\nu_t^\top$ is assumed to be positive definite. 
Note that the signal process is now a function of $\ep$, which is emphasized by a superscript "$(\ep)$".
As one can see, in the limit of $\ep\downarrow 0$, the system yields a free decoupled Gaussian signal process
for which the density function is exactly known. In the following, 
we try to expand the conditional density around it by taking into account 
state-dependent and observation effects perturbatively.

As explained in the previous section, we are interested in the unnormalized distribution
\bea
\rho^{(\ep)}(\psi(\xi))=\wt{\mbb{E}}\Bigl[\wt{Z}^{(\ep)}_t \psi(\xi,X_t^{(\ep)})\Bigr|\caly_t\Bigr] \nn
\eea
with
\bea
\wt{Z}_t^{(\ep)}=\exp\left(\ep \int_0^t H_s(X_s^{(\ep)})^\top dY_s-\frac{\ep^2}{2} \int_0^t
||H_s(X_s^{(\ep)})||^2 ds\right)~ \nn
\label{ztilde}
\eea
and
\bea
\left.\frac{d\mbb{P}}{d\wt{\mbb{P}}}\right|_{\calf_t}=\wt{Z}_t^{(\ep)}~. \nn
\eea

The corresponding Zakai equation becomes
\bea
\rho_t^{(\ep)}(\psi(\xi))=\rho_0(\psi(\xi))+\int_0^t \rho_s^{(\ep)}\Bigl(A_s^{(\ep)}\psi(\xi)\Bigr)ds
+\ep \int_0^t \rho_s^{(\ep)}\Bigl((H_s+B_s)^\top \psi(\xi)\Bigr)dY_s,
\label{ptd_zakai}
\eea
where conditional distribution $\rho$ is now also a function of $\ep$.
Here,
\bea
B_s^k=\sum_{i=1}^n(\gamma_s^\top(x))_{k,i}\frac{\part}{\part x_i}, \qquad k=1, \cdots, m. \nn
\eea
and the infinitesimal generator is given by
\bea
A_s^{(\ep)}&=&\sum_{i=1}^n\Bigl(f^i_s+\ep F^i_s(x)\Bigr)\frac{\part}{\part x_i}+\sum_{i,j=1}^n
\frac{1}{2}\Bigl(\nu_s+\ep \sigma_s(x)\Bigr)\Bigl(\nu_s+\ep \sigma_s(x)\Bigr)^\top _{ij}\frac{\part^2}{\part x_i \part x_j}\nn \\
&&+\sum_{i,j=1}^n\ep^2\frac{1}{2}\Bigl(\gamma_s(x)\gamma_s(x)^\top\Bigr)_{ij}\frac{\part^2}{\part x_i \part x_j}~. \nn
\eea
Our goal is to expand 
\bea
\rho_t^{(\ep)}(\psi(\xi))=\rho_t^{[0]}(\psi(\xi))+\ep \rho_t^{[1]}(\psi(\xi))+\ep^2 \rho_t^{[2]}(\psi(\xi))+\cdots
\label{asymp-expansion}
\eea
and obtain $\rho_t^{[j]}(\psi(\xi)),~j=\{0,1,2,\cdots\}$ up to an arbitrary order.
Here, we have defined
\bea
\rho_t^{[j]}(\psi(\xi)):=\left. \frac{1}{j !}\frac{\part^j}{\part \ep^j}\rho_t^{(\ep)}(\psi(\xi))\right|_{\ep=0}. \nn
\eea

\subsection{Recursive system for Asymptotic Expansion}
We now expand the Zakai equation for each order of $\ep$.
Note that, for any polynomial function $G$ of $x$,
one can write
\bea
G(x)e^{i\xi^\top x}=G(D_\xi)e^{i\xi^\top x} \nn
\eea
where, $G(D_\xi)$ denotes the differential operator obtained by replacing each $x_j$ in the function 
by $(D_\xi)_j$, which is a derivative operator defined as
\be
D_\xi=\frac{\part}{i\part \xi}~. \nn
\ee
This fact allows one to write
\bea
\rho_t^{(\ep)}(G\psi(\xi))=G(D_\xi)\rho_t^{(\ep)}(\psi(\xi)) \nn
\eea
which is linear for $\rho_t^{(\ep)}(\psi(\xi))$.

In order to avoid nonlinearity, we make use of this property of polynomial functions.
With slight abuse of notations, we treat $F_t(x), \sigma_t(x), \gamma_t(x)$
and $H_t(x)$ as arbitrary accurately approximated polynomial functions of $x$ (and time) for 
the corresponding original functions.
By Weierstrass' polynomial approximation theorem, this is always possible for any continuous function within 
a closed interval. In practice, one can take wide enough interval within which the signal process resides with 
probability sufficiently close to one and an associated polynomial approximation accurate enough for that range.

Then, one can formally write
\bea
A_s^{(\ep)}\psi(\xi,x)=\Bigl(A_s^{[0]}(\xi)+\ep A_s^{[1]}(\xi,D_\xi)+\ep^2A_t^{[2]}(\xi,D_\xi)\Bigr)\psi(\xi,x)
\label{inf_generator}
\eea
where
\bea
&&A_s^{[0]}(\xi)=i\xi^\top f_s-\frac{1}{2}\xi^\top (\nu_s\nu_s^\top)\xi  \nn \\
&&A_s^{[1]}(\xi,D_\xi)=i\xi^\top F_s(D_\xi)-\frac{1}{2}{\rm tr}\left[ \xi \xi^\top\Bigl(\nu_s\sigma_s^\top(D_\xi)+\sigma_s(D_\xi)\nu_s^\top
\Bigr) \right]\nn \\
&&A_s^{[2]}(\xi,D_\xi)=-\frac{1}{2}{\rm tr}\left[\xi \xi^\top\Bigl(\sigma_s(D_\xi)\sigma_s^\top(D_\xi)+
\gamma_s(D_\xi)\gamma_s^\top(D_\xi)\Bigr)\right] \nn
\eea
and similarly
\bea
(H_s(x)+B_s(x))^\top \psi(\xi,x)=\Bigl(H_s^\top(D_\xi)+i\xi^\top \gamma_s(D_\xi)\Bigr)\psi(\xi,x)~.
\label{obs_generator}
\eea
Substituting (\ref{asymp-expansion}), (\ref{inf_generator}) and (\ref{obs_generator}) into the Zakai 
equation (\ref{ptd_zakai}), 
one can easily confirm the following result.
\begin{theorem}
An arbitrary order of the asymptotic expansion $\rho_t^{[j]}(\psi(\xi))$, $j\in\{0,1,2,\cdots\}$ 
\be
\rho_t^{(\ep)}(\psi(\xi))=\rho_t^{[0]}(\psi(\xi))+\ep \rho_t^{[1]}(\psi(\xi))+\ep^2 \rho_t^{[2]}(\psi(\xi))+\cdots
\ee
of the unnormalized filtered characteristic function 
\be
\rho_t^{(\ep)}(\psi(\xi))=\wt{\mbb{E}}\Bigl[\exp\Bigl(i\xi^\top X_t^{(\ep)}\Bigr)\wt{Z}_t^{(\ep)}\Bigr|\caly_t\Bigr] \nn
\ee
satisfies 
\bea
d\rho_t^{[j]}(\psi(\xi))&=&A_t^{[0]}(\xi)\rho_t^{[j]}(\psi(\xi))dt\nn \\
&&+\Bigl\{A_t^{[1]}(\xi,D_\xi)\rho_t^{[j-1]}(\psi(\xi))+A_t^{[2]}(\xi,D_\xi)\rho_t^{[j-2]}(\psi(\xi))\Bigr\}dt\nn \\
&&+\Bigl(H_t^\top(D_\xi)+i\xi^\top \gamma_t(D_\xi)\Bigr)\rho_t^{[j-1]}(\psi(\xi))dY_t 
\label{zakai-asymp}
\eea
with initial conditions $\rho_0^{[0]}(\psi(\xi))=\rho_0(\psi(\xi)),~\rho_0^{[j]}(\psi(\xi))=0~(j\geq 1)$ and the convention that 
\bea
\rho^{[k]}(\psi(\xi))\equiv 0 \nn
\eea
for $k<0$.
\label{theorem1}
\end{theorem}

Considering a special case where there is no observation, one obtains a simple corollary for a
standard unconditional characteristic function.
\begin{corollary}
An arbitrary order of the asymptotic expansion $\rho_t^{[j]}(\psi(\xi))$, $j\in\{0,1,2,\cdots\}$ 
\bea
\rho_t^{(\ep)}(\psi(\xi))=\rho_t^{[0]}(\psi(\xi))+\ep \rho_t^{[1]}(\psi(\xi))+\ep^2 \rho_t^{[2]}(\psi(\xi))+\cdots
\eea
of the characteristic function
\be
\rho_t^{(\ep)}(\psi(\xi))=\mbb{E}\Bigl[\exp\Bigl(i\xi^\top X_t^{(\ep)}\Bigr)\Bigr] \nn
\ee
satisfies
\bea
d\rho_t^{[j]}(\psi(\xi))&=&A_t^{[0]}(\xi)\rho_t^{[j]}(\psi(\xi))dt\nn \\
&&+\Bigl\{A_t^{[1]}(\xi,D_\xi)\rho_t^{[j-1]}(\psi(\xi))+A_t^{(2)}(\xi,D_\xi)\rho_t^{[j-2]}(\psi(\xi))\Bigr\}dt\nn
\eea
with initial conditions $\rho_0^{[0]}(\psi(\xi))=\rho_0(\psi(\xi)),~\rho_0^{[j]}(\psi(\xi))=0~(j\geq 1)$
and the convention that 
\bea
\rho^{[k]}(\psi(\xi))\equiv 0 \nn
\eea
for $k<0$.
\label{corollary1}
\end{corollary}


The above result shows that one only has to deal with a set of decoupled ODEs in terms of momentum $\{\xi\}$ with 
a given observation path of $Y$.  It is straightforward to solve the above equation for each $\xi$
up to a certain $\ep$-order, and use discrete Fourier transformation technique to obtain the density function.
In Fourier analysis of smooth functions, it is well-known that most of the information is carried by small number of modes.
In fact, in an example we provide in a later section, the resultant density function does not change meaningfully 
once the number of $\xi$-mode reaches $\sim 30$.
This feature combined with the decoupled dynamics of characteristic function is expected to weaken
the curse of dimensionality significantly, at least compared to typical PDE approaches.

Analytical calculation is also possible. Since the dynamics is linear, 
one easily obtains the following results:\\\\
{\bf{Zeroth order}}
\bea
\rho_t^{[0]}(\psi(\xi))=e^{\int_0^t A_s^{[0]}(\xi)ds}\rho_0(\psi(\xi)) 
\label{rho-0th}
\eea
{\bf{First order}}
\bea
\rho_t^{[1]}(\psi(\xi))&=&\int_0^t e^{\int_s^t A_u^{[0]}(\xi)du}\Bigl\{
A_s^{[1]}(\xi,D_\xi)\rho_s^{[0]}(\psi(\xi))ds \nn \\
&&\qquad+\Bigl(H_s^\top(D_\xi)+i\xi^\top \gamma_s(D_\xi)\Bigr)\rho_s^{[0]}(\psi(\xi))dY_s\Bigr\}
\label{rho-1st}
\eea
{\bf{Higher order $(j\geq 2)$}}\\
Using the lower order results, an arbitrary order of the expansion can be expressed recursively as 
\bea
\rho_t^{[j]}(\psi(\xi))&=&\int_0^t e^{\int_s^t A_u^{[0]}(\xi)du}\Bigl\{
\Bigl(A_s^{[1]}(\xi,D_\xi)\rho_s^{[j-1]}(\psi(\xi))+A_s^{[2]}(\xi,D_\xi)\rho_s^{[j-2]}(\psi(\xi))\Bigr)ds\nn \\
&&\quad+\Bigl(H_s^\top(D_\xi)+i\xi^\top \gamma_s(D_\xi)\Bigr)\rho_s^{[j-1]}(\psi(\xi))dY_s\Bigr\}~. \nn \\
\eea

Rigorous mathematical proofs of the validity and convergence of the above asymptotic expansion when taking
$\ep \downarrow 0$ are beyond the
scope of the current work. However, for a given observation path $Y$, it is likely to be 
proved by  a similar line of arguments for the asymptotic expansion in position space without filtering problem 
given in Takahashi (1999), which is based on the results of Yoshida (1992a, 1992b, 1997) and Ikeda and Watanabe (1989).
In the next section, we explain the inversion method to obtain the density function.

\subsection{Density Formula}
\label{sec-density}
In this section, we provide a strategy to obtain an analytical expression of the 
filtered density. Although this is not necessary if one is only interested in numerical implementation
with discrete Fourier transformation, the analytical expression can be quite useful for 
various applications in finance. In particular, a model calibration and quick response to a client request of 
price indication require very fast evaluation.

Let us consider the inverse Fourier transformation of $\rho_t^{(\ep)}(\psi(\xi))$:
\bea
\phi_t^{(\ep)}(z)=\frac{1}{(2\pi)^n}\int_{\mbb{R}^n}
e^{-i\xi^\top z}\rho_t^{(\ep)}(\psi(\xi))d^n\xi~. \nn
\eea
This corresponds to the unnormalized conditional probability density of the signal $X^{(\ep)}_t$ 
given the observation path of $Y$ (See the discussion in Sec.\ref{c-function}.).
The desired normalized conditional probability density of the 
signal is then given by
\bea
\bar{\phi}_t^{(\ep)}(z)=\frac{1}{c_t^{(\ep)}}\phi_t^{(\ep)}(z)  \nn
\eea
where $c_t^{(\ep)}$ is a normalization factor defined as
\bea
\Bigl(c_t^{(\ep)}\Bigr)^{-1}=\int_{\mbb{R}^n}\phi_t^{(\ep)}(z)d^n z~. \nn
\eea
Thus, for applications, it suffices to calculate the expression of $\phi_t^{(\ep)}(z)$.

\subsubsection{Gaussian distribution for $X_0$}
For simplicity, let us first consider the case where  $X_0$ is distributed by a Gaussian law 
$\caln(x_0;\Sigma_0)$ with mean $x_0$ and 
the covariance $\Sigma_0$ of a symmetric positive definite matrix.
In this case, we have
\bea
\rho_0(\psi(\xi))&=&\int_{\mbb{R}^n}e^{i\xi^\top x}n[x;x_0,\Sigma_0]d^n x  \nn \\
&=&\frac{1}{\sqrt{(2\pi)^n|\Sigma_0|}}\int_{\mbb{R}^n}e^{i\xi^\top x}  
\exp\left(-\frac{1}{2}(x-x_0)^\top \Sigma_0^{-1}(x-x_0)\right)d^nx \nn
\eea
where $n[x;x_0,\Sigma_0]$ is the probability density function for a random variable with 
Gaussian low of $\caln(x_0;\Sigma_0)$, and $|\Sigma_0|$ denotes the determinant of $\Sigma_0$.
The evaluation can be done easily by considering the variable change from $x$ to $\eta$ given by
\be
x=x_0+P_0\eta \nn
\ee
with a matrix $P_0$ satisfying 
\be
\Sigma_0=P_0P_0^\top. \nn
\ee
Integration in terms of $\eta$ leads to 
\bea
\rho_0(\psi(\xi))=\exp\Bigl(i\xi^\top x_0-\frac{1}{2}\xi^\top \Sigma_0\xi\Bigr)~. \nn
\eea

Then, from (\ref{rho-0th}) of the previous section, we have
\bea
\rho_t^{[0]}(\psi(\xi))=\exp\Bigl(i\xi^\top x_t-\frac{1}{2}\xi^\top \Sigma_t \xi\Bigr)
\label{rho-t-0}
\eea
where
\bea
&&x_t=x_0+\int_0^t f_s ds \nn \\
&&\Sigma_t=\Sigma_0+\int_0^t\nu_s\nu_s^\top ds~. \nn
\eea
Thus, it is clear that $X_t^{[0]}$ has a Gaussian distribution $\caln(x_t;\Sigma_t)$.
If the initial position of  $X_0$ is exactly known as $X_0=x_0$, then one clearly has
\be
\rho_0(\psi(\xi))=e^{i\xi^\top x_0} \nn
\ee
and hence one can simply insert $\Sigma_0=0$ in (\ref{rho-t-0}).

By the property of the exponential form and $(\ref{rho-1st})$, one can check that $\rho_t^{[1]}(\psi(\xi))$ 
is given by
\bea
\rho_t^{[1]}(\psi(\xi))=\rho_t^{[0]}(\psi(\xi))\left(\int_0^t a_s(\xi)ds+b_s(\xi)dY_s\right) 
\label{rho-t-1}
\eea
with polynomial functions $a_s(\xi)\in\mbb{R}$ and $b_s(\xi)\in\mbb{R}^{1\times m}$ of $\xi$ 
\bea
&&a_s(\xi)=\rho_s^{[0]}(\psi(\xi))^{-1}A_s^{(1)}(\xi,D_\xi)\rho_s^{[0]}(\psi(\xi)) \nn \\
&&b_s(\xi)=\rho_s^{[0]}(\psi(\xi))^{-1}\Bigl(H_s(D_\xi)^\top+i\xi^\top \gamma_s(D_\xi)\Bigr)\rho_s^{[0]}(\psi(\xi)) \nn
\eea
which can be expressed by Hermite polynomials in general.

Thus, the first order correction to the unnormalized conditional density can be expressed as
\bea
\phi_t^{[1]}(z)&:=&\frac{1}{(2\pi)^n}\int_{\mbb{R}^n}e^{-i\xi^\top z}\rho_t^{[1]}(\psi(\xi))d^n\xi \nn \\
&=&\left(\int_0^t a_s(D_z)ds+b_s(D_z)^\top dY_s\right)n[z;x_t,\Sigma_t]~. 
\label{1st-density}
\eea
Here,
\bea
D_z:=i\frac{\part}{\part z} \nn
\eea
and $a_s(D_z)$, $b_s(D_z)$ denote the derivative operator of $z$ obtained by replacing each $\xi$ in the functions
by $D_z$. 

Repeating the same arguments, one can see that $\rho_t^{[j]}(\psi(\xi))$ can be given, as a generalization of (\ref{rho-t-1}), by
\bea
\rho_t^{[j]}(\psi(\xi))&=&\rho_t^{[0]}(\psi(\xi))\int_0^t\int_0^{s_{j}}\cdots\int_0^{s_2}\Bigl(\Gamma_{s_1,\cdots,s_j}(\xi)ds_1ds_2\cdots ds_j\nn \\
&&\hspace{20mm}+\Gamma_{y_1,s_2,\cdots,s_j}(\xi)~dY_{s_1}ds_2\cdots ds_j\nn \\
&&\qquad\hspace{20mm}\cdots\nn \\
&&\hspace{20mm}+\Gamma_{y_1,y_2,\cdots,y_j}(\xi)~dY_{s_1}dY_{s_2}\cdots dY_{s_j}\Bigr)+\cdots 
\label{FT-density}
\eea
with certain polynomial functions $\{\Gamma_{s_1,\cdots,s_j}(\xi),\cdots,\Gamma_{y_1,\cdots,y_j}(\xi)\}$ 
of $\xi$ with appropriate dimensions. The $\{\cdots\}$ denotes the term with integration order of 
$(<j)$, which stems from the existence of the $\ep$-second order operator $A^{[2]}$.
Then, as a generalization of (\ref{1st-density}), the $j$-th order term of the unnormalized conditional density is also 
given by the correction to the Gaussian distribution:
\bea
\phi_t^{[j]}(z)&:=&\frac{1}{(2\pi)^n}\int_{\mbb{R}^n}e^{-i\xi^\top z}\rho_t^{[j]}(\psi(\xi))d^n\xi \nn \\
&=&\int_0^t\int_0^{s_{j}}\cdots\int_0^{s_2}\Bigl(\Gamma_{s_1,\cdots,s_j}(D_z)ds_1ds_2\cdots ds_j\nn \\
&&+\Gamma_{y_1,s_2,\cdots,s_j}(D_z)~dY_{s_1}ds_2\cdots ds_j\nn \\
&&\hspace{20mm}\cdots\nn \\
&&+\Gamma_{y_1,y_2,\cdots,y_j}(D_z)~dY_{s_1}dY_{s_2}\cdots dY_{s_j}\Bigr)n[z;x_t,\Sigma_t]~+\cdots. 
\label{inverse-form}
\eea
In the case of no (or trivial) observation, one can get the asymptotic expansion of unconditional 
probability density by putting $dY$ terms zero.

\subsubsection{Non-Gaussian distribution for $X_0$}
Even when the initial distribution is not exactly Gaussian, if one can approximate it by
the form
\bea
\phi_0(z)=({\mbox{some polynomial function of z}})\times n[z;x_0,\Sigma_0]~, 
\label{gram-charlier}
\eea
then the properties of the inverse transformation 
given in the previous section still hold in almost the same way.
This is, for example,  the case when one approximates the initial distribution by Gram-Charlier expansions.
In the case when (\ref{gram-charlier}) holds, one can still write $\rho_t^{[0]}$ in the form
\bea
\rho_t^{[0]}(\psi(\xi))= ({\mbox{some polynomial function of }}\xi)\times \exp\Bigl(i\xi^\top x_t-\frac{1}{2}
\xi^\top \Sigma_t \xi\Bigr) \nn
\eea
and it only changes the functions $\{\Gamma\}$ in (\ref{FT-density}) and (\ref{inverse-form}).

\section{A Direct Application to Kushner-Stratonovich Equation}
We can also apply the technique to the Kushner-Stratonovich (KS) equation
that describes the dynamics of the normalized conditional density of $\pi_t$
instead of $\rho_t$. Although it suffices to work on the simpler Zakai equation in filtering problems, we
directly treat KS equation  here 
to demonstrate the fact that the asymptotic expansion can also be applied to 
measure-valued non-linear systems.
For the setup given in Sec.\ref{sec-zakai}, the Kushner-Stratonovich equation is
given by
\bea
d\pi_t(\varphi)=\pi_t(A_t\varphi)dt+
\Bigl\{\pi_t\Bigl( (h_t+B_t)^\top\varphi\Bigr)-\pi_t(h_t^\top)\pi_t(\varphi)\Bigr\}(dY_t-\pi_t(h_t)dt) \nn
\eea
with a given initial value $\pi_0(\varphi)$. This is clearly a nonlinear equation for the 
measure-valued process $\pi_t$. See a textbook Bain and Crisan (2008) for details of the derivation.

Let us now introduce the same perturbed system as in Sec.\ref{sec-perturbed}.
Then, one obtains the KS equation for $\psi(\xi,\cdot)$ as
\bea
&&d\pi_t^{(\ep)}(\psi(\xi))=\pi_t^{(\ep)}(A_t^{(\ep)}\psi(\xi))dt\nn \\
&&\qquad+\ep\Bigl\{\pi_t^{(\ep)}\Bigl((H_t+B_t)^\top \psi(\xi)\Bigr)-\pi_t^{(\ep)}(H_t^\top)\pi_t^{(\ep)}(\psi(\xi))\Bigr\}
(dY_t-\ep \pi_t^{(\ep)}(H_t)dt)\nn
\eea
By the same polynomial-function approximations, one can rewrite it as
\bea
&&d\pi_t^{(\ep)}(\psi(\xi))=\Bigl(A_t^{[0]}(\xi)+\ep A_t^{[1]}(\xi,D_\xi)+\ep^2 A_t^{[2]}(\xi,D_\xi)\Bigr)\pi_t^{(\ep)}(\psi(\xi))dt\nn\\
&&\qquad+\ep\Bigl\{ \Bigl(H_t^\top (D_\xi)+i\xi^\top \gamma_t(D_\xi)\Bigr)\pi_t^{(\ep)}(\psi(\xi))-\pi_t^{(\ep)}(H_t^\top)\pi_t^{(\ep)}(\psi(\xi))\Bigr\}
(dY_t-\ep \pi_t^{(\ep)}(H_t)dt)~.\nn\\ 
\label{ks_eq}
\eea
As before, we try to expand the solution as
\be
\pi_t^{(\ep)}(\psi(\xi))=\pi_t^{[0]}(\psi(\xi))+\ep\pi_t^{[1]}(\psi(\xi))+\ep^2 \pi_t^{[2]}(\psi(\xi))+\cdots
\ee
with the definition 
\bea
\pi_t^{[j]}(\psi(\xi)):=\left.\frac{1}{j!}\frac{\part^j}{\part \ep^j}\pi_t^{(\ep)}(\psi(\xi))\right|_{\ep=0}~.
\eea
Note that there appears $\pi_t^{(\ep)}(H_t)$ in (\ref{ks_eq}). This term does no harm since an arbitrary order $j$ of asymptotic expansion, we need $\pi_t^{[i]}(H_t)$ only for $i=0,1,\cdots, j-1$ due to the additional $\ep$-factor.
Thus, at the calculation of $j$-th order expansion, one can use
\be
\pi_t^{[i]}(H_t)=H_t(D_\xi)\pi_t^{[i]}(\psi(\xi))\Bigr|_{\xi=0} \nn
\ee
where $\pi_t^{[i]}(\psi(\xi))$ are already known for $i\leq j-1$.
\\
\\
Let us give the first few orders of expansions for the KS equation:\\
{\bf{Zeroth order}}
\bea
d\pi_t^{[0]}(\psi(\xi))=A_t^{[0]}(\xi)\pi_t^{[0]}(\psi(\xi))dt
\eea
with initial value $\pi_0^{[0]}(\psi(\xi))=\pi_0(\psi(\xi))$.\\
{\bf{First oder}}
\bea
&&d\pi_t^{[1]}(\psi(\xi))=A_t^{[0]}(\xi)\pi_t^{[1]}(\psi(\xi))dt\nn \\
&&\qquad+\Bigl\{\Bigl(H_t^\top(D_\xi)+i\xi^\top\gamma_t(D_\xi)\Bigr)\pi_t^{[0]}(\psi(\xi))-
\pi_t^{[0]}(H_t^\top)\pi_t^{[0]}(\psi(\xi))\Bigr\}dY_t
\eea
with $\pi_0^{[1]}(\psi(\xi))=0$. \\
{\bf{Second order}}
\bea
&&d\pi_t^{[2]}(\psi(\xi))=A_t^{[0]}(\xi,D_\xi)\pi_t^{[2]}(\psi(\xi))dt\nn \\
&&\qquad+\Bigl\{A_t^{[1]}(\xi,D_\xi)\pi_t^{[1]}(\psi(\xi))+A_t^{[2]}(\xi,D_\xi)\pi_t^{[0]}(\psi(\xi))\nn \\
&&\qquad~-\Bigl(H_t^\top(D_\xi)+i\xi^\top\gamma_t(D_\xi)-\pi_t^{[0]}(H_t^\top)\Bigr)\pi_t^{[0]}(\psi(\xi))\pi_t^{[0]}(H_t)
\Bigr\}dt\nn \\
&&\qquad+\Bigl\{\Bigl(H_t^\top(D_\xi)+i\xi^\top\gamma_t(D_\xi)-\pi_t^{[0]}(H_t^\top)\Bigr)\pi_t^{[1]}(\psi(\xi))
-\pi_t^{[1]}(H_t^\top)\pi_t^{[0]}(\psi(\xi))\Bigr\}dY_t\nn\\
\eea
with initial condition $\pi_t^{[2]}(\psi(\xi))=0$.\\\\
Although the system of ODEs does not keep the same structure as the unnormalized distribution $\rho$, 
it is clear that one can still perform the perturbation 
order by order. Furthermore, from the discussion given in the next section,
it is not always necessary to derive higher order asymptotic expansion for accurate estimation.

\section{Substepping Method for Asymptotic Expansion}
\label{substep}
It is obvious by construction that
the accuracy of approximation deteriorates once the cumulative contributions from 
perturbation terms
\bea
\ep F, \quad \ep \sigma, \quad \ep \gamma, \quad \ep H \nn
\eea
become significant. This is a common problem of asymptotic expansion methods for various applications. 
In particular for the filtering problems, requiring small perturbation terms seems rather restrictive
since it indicates that one can treat only noisy observations (i.e. small $h$).
In financial applications of the position-space asymptotic expansion, it is known that one needs higher order 
approximation to reach enough accuracy for practical use in long-dated or high-volatility 
environments. There exist many efforts to obtain higher order corrections
systematically to tackle these problems. See, for example, Takahashi and Takehara (2007), Takahashi et al. (2010),
and Li (2010) for 
recent developments in this direction.

Let us consider the problem in the momentum-space approach.
In Theorem~\ref{theorem1}, we have seen that the equation $(\ref{zakai-asymp})$
determines the correction terms with a given initial condition $\rho_0(\psi(\xi))$.
Although higher order calculation is straightforward, there exists a simpler and more efficient 
way to improve the approximation. 
An obvious but important feature of (\ref{zakai-asymp}) is that the recursion can be started from 
an arbitrary initial distribution $\rho_0(\psi(\xi))$. Since asymptotic expansion generally
works very well for short maturities,  the above feature 
naturally leads to the following substepping idea for asymptotic expansion.\\
\\
{\it
(1) Create an appropriate time grid $\{0=T_0<T_1<\cdots<T_N=t\}$
in such a way that the asymptotic expansion converges well within each sub-period $[T_{i-1},T_i]$.\\
(2) Solve (\ref{zakai-asymp}) of each $\xi$ for $s\in[T_0,T_1]$ up to the $k$-th order of the asymptotic 
expansion. This can be the second (or even the first) order if the stepping size is small enough. \\
(3) Update the initial condition for the next period $[T_1,T_2]$ by setting
\be
\rho_{T_1}^{[0]}(\psi(\xi))\leftarrow \left(\sum_{j=0}^k\ep^j\rho_{T_1}^{[j]}(\psi(\xi))\right){\mbox{obtained in step(2)}} ~.
\ee
(4) Solve (\ref{zakai-asymp}) for $[T_1,T_2]$ with the updated initial condition.\\
(5) Repeat the procedures till the final period to obtain $\rho_{T_N}^{[j]}(\psi(\xi))$.
}
\\
\\
Although performing the above method analytically by hand is quite laborious
due to a large number of derivative operations,
one can do it quite efficiently in numerical implementations. 
This is because amount of procedures required for $dt$ and $dY$
integration does not change in the above operations. 
In fact, one obtains accurate results faster by performing finer substepping with lower-order approximation
than performing higher-order approximation without substepping.

The substepping method may be  very useful for general measure-valued nonlinear equations.
Although it is tedious to obtain the asymptotic expansion for complicated dynamics, it is definitely possible 
for the first few orders as we have demonstrated by using the Kushner-Stratonovich equation.
 If the approximation works well, at least within a very short period, the above numerical procedures will extend the 
effective region for the asymptotic expansion.
If it is applied to a standard unconditional characteristic function,
it should also offer an efficient option pricing method, particularly for long-dated
and high-volatile setups. 
\section{Examples}
\subsection{Analytical Application to CIR Process}
Let us first consider the approximation of one-dimensional CIR process with no filtering issue,
which helps to obtain a concrete image how analytical procedures work. We study
\be
dX_t=\theta(\mu-X_t)dt+\sigma\sqrt{X_t}dV_t \nn
\ee
with $X_0=\mu$. All the parameters $\theta, \mu$ and $\sigma$ are positive constants
satisfying $2\theta\mu>\sigma^2$. Then, the probability density of $X_t$ is known to have
a non-central chi-squared distribution.

For asymptotic expansion, we treat it as the following perturbed system:~\footnote{One needs to put $\ep=1$
at the end for the comparison to the original model.}
\bea
dX_t^{(\ep)}=\ep \theta F(X_t^{(\ep)})dt+\sigma\Bigl(\sqrt{\mu}+\ep R(X_t^{(\ep)})\Bigr)dV_t \nn
\eea
with $X_0^{(\ep)}=\mu$. Here, we have defined
\bea
&&F(x)=\mu-x \nn \\
&&R(x)={\mbox{Taylor expansion at $(x=\mu)$ of } }(\sqrt{x}-\sqrt{\mu})~. \nn
\eea
In this example, we are going to adopt the $3$rd-order expansion for $R(x)$.
Note that, Taylor expansion provides a good polynomial approximation only when 
the process $X$ resides near $\mu$. If volatility is very high, it may be better
to perform a different method, such as the minimization of the least square difference
for appropriate range. We shall see some examples in the next section.
Systematic strategy for determining the optimal choice of polynomial function
remains as an important future work.

Then, the infinitesimal generator is given by
\bea
A^{(\ep)}=\ep \theta F(x)\frac{\part}{\part x}+\frac{1}{2}\sigma^2\Bigl(\sqrt{\mu}+\ep R(x)\Bigr)^2
\frac{\part^2}{\part x^2} \nn
\eea
and hence
\bea
&&A^{[0]}(\xi)=-\frac{1}{2}\xi^2\sigma^2\mu  \nn \\
&&A^{[1]}(\xi,D_\xi)=i\xi \theta F(D_\xi)-\xi^2\sigma^2\sqrt{\mu}R(D_\xi)  \nn \\
&&A^{[2]}(\xi,D_\xi)=-\frac{1}{2}\xi^2\sigma^2 R^2(D_\xi)~. \nn
\eea
From Corollary~\ref{corollary1}, analytical calculation can be performed as follows: 
\subsubsection{Zeroth order}
We have
\bea
d\rho_t^{[0]}(\psi(\xi))=A^{[0]}(\xi)\rho_t^{[0]}(\psi(\xi))dt \nn 
\eea
with $\rho_0^{[0]}(\psi(\xi))=e^{i\xi \mu}$.
Thus, it gives
\bea
\rho_t^{[0]}(\psi(\xi))=\exp\left(i\xi \mu-\frac{1}{2}\xi^2 \sigma^2 \mu t\right)~. \nn
\eea
Then, one obtains the zeroth order density function as
\bea
\phi_t^{[0]}(z)=\frac{1}{\sqrt{2\pi\Sigma_t}}\exp\left(-\frac{(\mu-z)^2}{2\Sigma_t}\right) \nn
\eea
with the definition of $\Sigma_t=\mu\sigma^2 t$.

\subsubsection{First order}
The first order correction is given by
\bea
\rho_t^{[1]}(\psi(\xi))=\int_0^t e^{(t-s)A^{[0]}(\xi)}A^{[1]}(\xi,D_\xi)\rho_s^{[0]}(\psi(\xi))ds~. \nn
\eea
By straightforward differential operations lead to
\bea
\Bigl(\rho_s^{[0]}(\psi(\xi))\Bigr)^{-1}A^{[1]}(\xi,D_\xi)\rho_s^{[0]}(\psi(\xi))&=&\frac{1}{8}s\sigma^2(8\theta\mu+\sigma^2)\xi^2
-\frac{1}{16}i s \sigma^4(8\mu+3 s\sigma^2)\xi^3\nn \\
&&-\frac{1}{8}s^2\mu \sigma^6 \xi^4+\frac{1}{16}is^3 \mu \sigma^8 \xi^5~. \nn
\eea
Then, following the procedures of Sec.~\ref{sec-density}, one obtains
\bea
\rho_t^{[1]}(\psi(\xi))=\rho_t^{[0]}\Bigl(a_2 \xi^2 +a_3 \xi^3 +a_4 \xi^4 +a_5 \xi^5\Bigr) \nn
\eea
and also
\bea
\phi_t^{[1]}(z)&=&\frac{1}{2\pi}\int_{\mbb{R}}e^{-i\xi z}\rho_t^{[1]}(\psi(\xi))d\xi \nn \\
&=&\left(-a_2\frac{\part}{\part z^2}-ia_3\frac{\part^3}{\part z^3}+a_4\frac{\part^4}{\part z^4}
+ia_5\frac{\part^5}{\part z^5}\right)\phi_t^{[0]}(z) \nn
\eea
with the coefficients defined by
\bea
&&a_2= \frac{1}{16}t^2\sigma^2(8\theta \mu+\sigma^2),\qquad a_3= -\frac{1}{16}it^2\sigma^4(4\mu+t\sigma^2) \nn \\
&&a_4=-\frac{1}{24}t^3\mu \sigma^6,\qquad a_5=\frac{1}{64}i t^4 \mu \sigma^8~. \nn
\eea

\begin{figure}
\begin{center}	
\includegraphics[width=74mm]{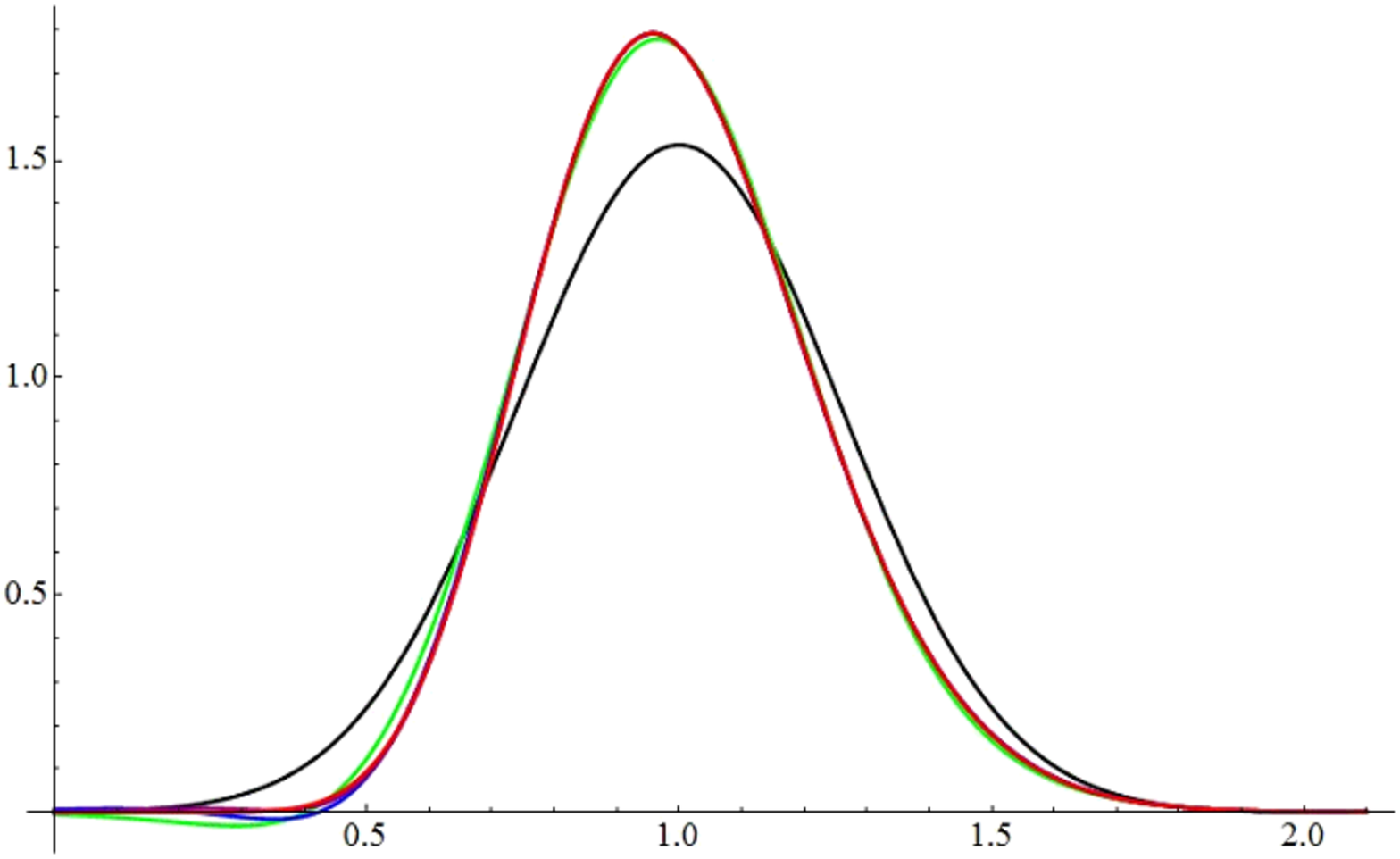} 
\includegraphics[width=74mm]{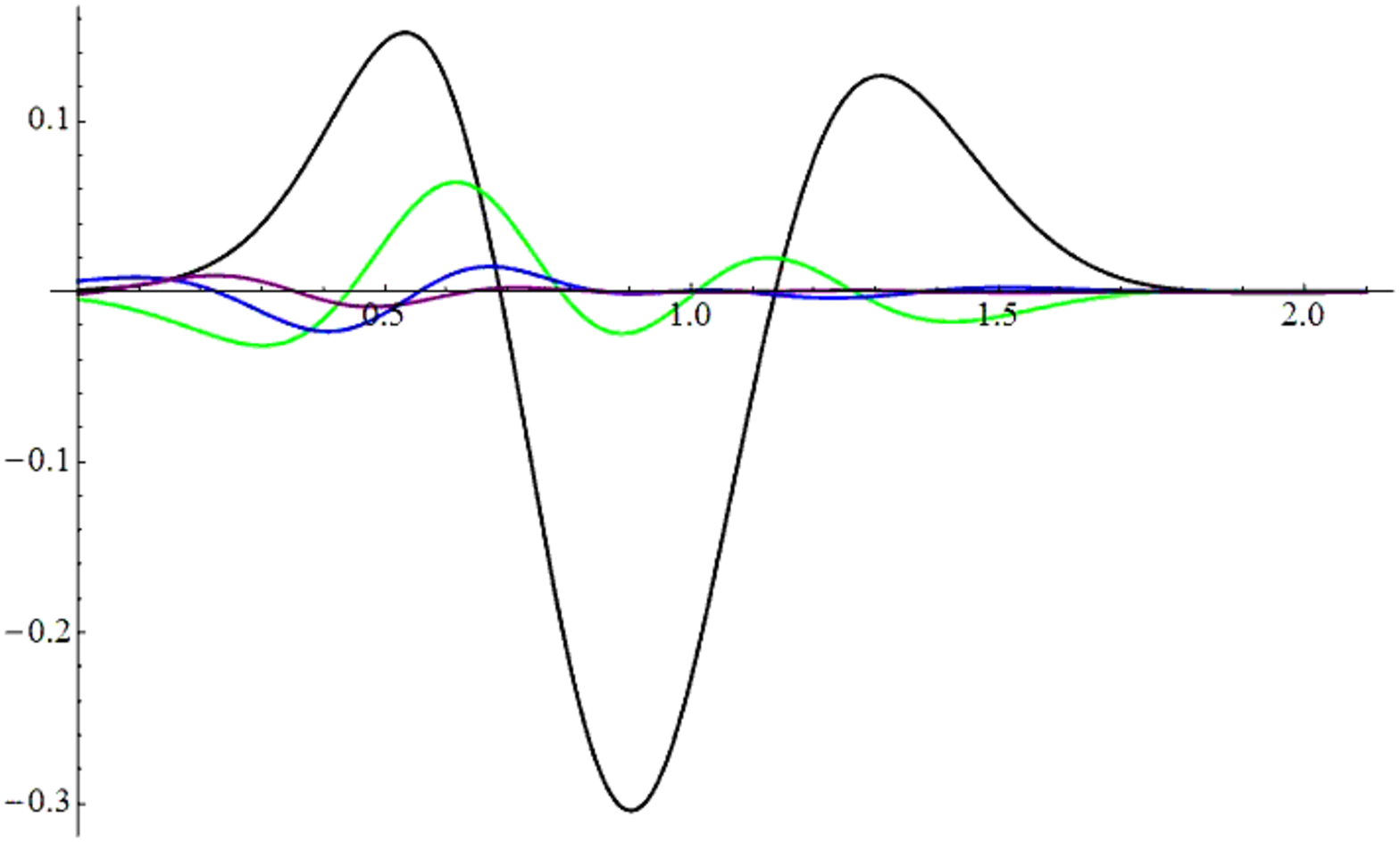} 
\end{center}
\caption{$t=3, \mu=1, \sigma=0.15, \theta=0.1$. 
(Black, Green, Blue, Purple) lines denote (0th, 1st, 2nd, 3rd) order approximation of asymptotic expansion, respectively.
Red line denotes the exact density function given by a non-central chi-squared distribution.
The second graph represents the difference from the exact density function.}
\label{fig1}
\end{figure}

\begin{figure}
\begin{center}	
\includegraphics[width=74mm]{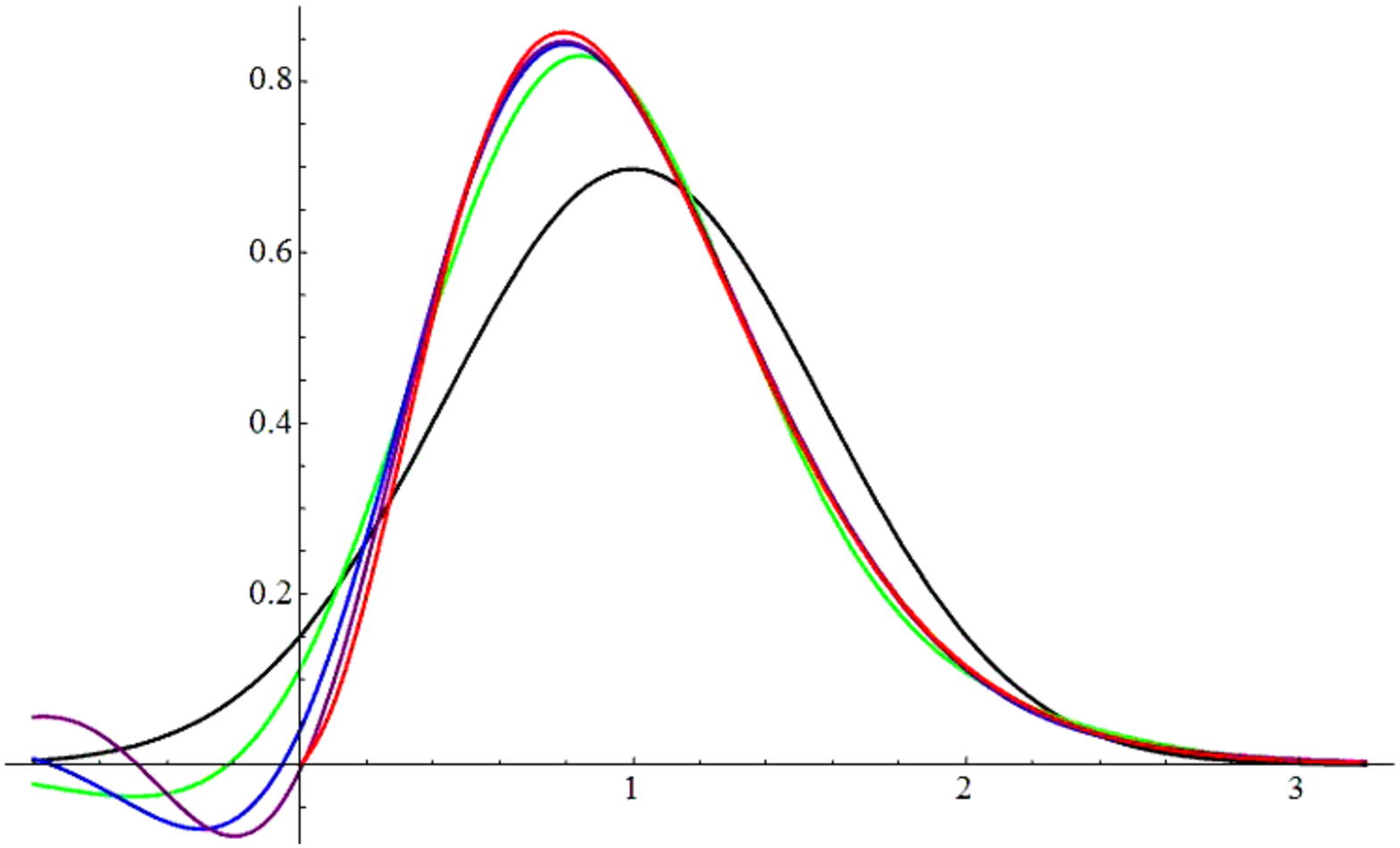}
\includegraphics[width=74mm]{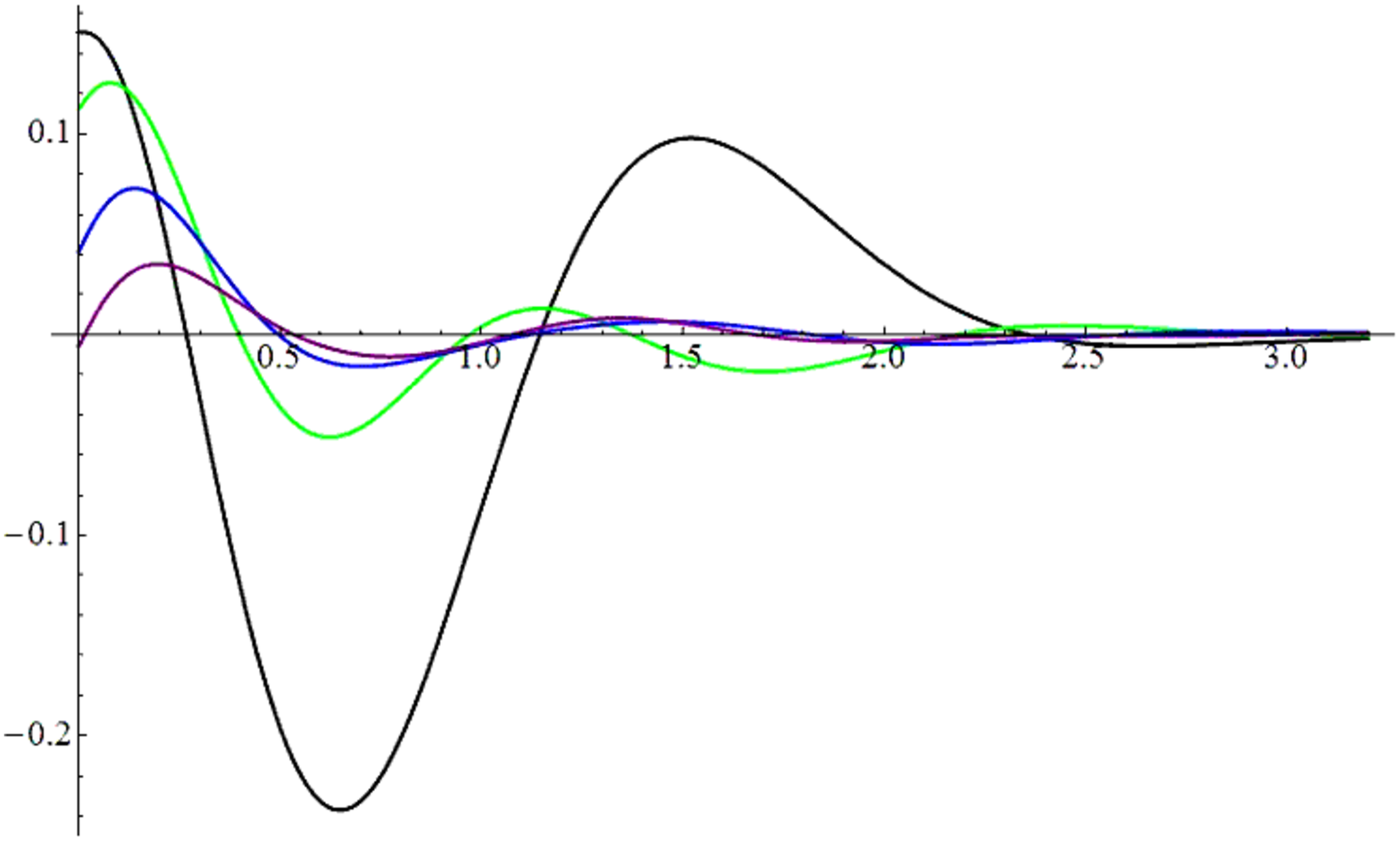} 
\end{center}
\caption{$t=3, \mu=1, \sigma=0.33, \theta=0.1$.
 (Black, Green, Blue, Purple) lines denote (0th, 1st, 2nd, 3rd) order approximation of asymptotic expansion, respectively.
Red line denotes the exact density function given by a non-central chi-squared distribution. The second graph represents the difference from the exact density function.}
\label{fig2}
\end{figure}

Higher order expressions follow similarly with the help of analytical software if necessary.
In Figs.~\ref{fig1} and \ref{fig2}, we have given numerical results up to $\ep$-3rd order asymptotic expansions
without substepping compared with the exact non-central chi-squared distribution.
When volatility is large, there appears sizable deviation from the correct distribution for small $x$.
This is understandable because Taylor expansion near the origin is not accurate. 
Except a neighbor of the origin and $(x<0)$,  
one can see that our approximation reproduces the desired density function well.


\subsection{A Numerical Application to  Bene$\check{\mbox{s}}$ Filter}
Next, as the second example, we study the Bene$\check{\mbox{s}}$ filter [Bene$\check{\mbox{s}}$ (1981)], where the drift of the signal process is nonlinear.
This is a special case for which there exists an exact solution in the class of non-Gaussian filtering problems,
thus it is quite useful for testing the current method.
In the class of Bene$\check{\mbox{s}}$ filter, we choose a following one-dimensional example:
\bea
&&dX_t=f(X_t)dt+\sigma dV_t \nn \\
&&dY_t=\Bigl(h_1 X_t+h_2\Bigr)dt+dW_t \nn
\eea
with $X_0=Y_0=0$, where $f(x)$ is given by 
\be
f(x)=a\sigma \tanh\Bigl(a \frac{x}{\sigma}\Bigr)~. \nn
\ee
Here, $a,\sigma, h_1$ and $h_2$ are all constants.
In this case, the exact filtered density of $X_t$ is given by
\bea
&&\pi_t^{\rm exact}(z)\nn \\
&&\quad=\frac{1}{n_t}\cosh\Bigl(a\frac{z}{\sigma}\Bigr)\exp\left(
-\frac{h_1}{2\sigma}\coth(th_1\sigma)z^2\nn \right. \\
&&\hspace{20mm}\left.+\Bigl(
h_1\int_0^t \frac{\sinh(s h_1\sigma)}{\sinh(t h_1\sigma)}dY_s+
\frac{h_2}{\sigma\sinh(th_1\sigma)}-\frac{h_2}{\sigma}\coth(th_1\sigma)\Bigr)z \right)\nn 
\eea 
where $n_t$ is the normalization factor to guarantee
\be
\int_{\mbb{R}}\pi_t^{\rm{exact}}(z)dz=1~. \nn
\ee

For this problem, we setup the following perturbed approximation:~\footnote{As before, one needs to put $\ep=1$ at the
end for the comparison to the original model.}
\bea
&&dX_t^{(\ep)}=\ep F(X_t^{(\ep)})dt+\sigma dV_t \nn \\
&&dY_t=\ep H(X_t^{(\ep)})dt+dW_t \nn
\eea
with $X_0^{(\ep)}=Y_0=0$.
Here, we use
\bea
&&H(x)=h_1 x+h_2 \nn \\
&&F(x)={\mbox{(polynomial approximation of) }}f(x)~. \nn
\eea
We explain the details of polynomial approximation later.


The infinitesimal generator contains only up to $\ep$-first order term. We have
\bea
A^{(\ep)}=\ep F(x)\frac{\part}{\part x}+\frac{1}{2}\sigma^2 \frac{\part^2}{\part x^2} \nn
\eea
and 
\bea
A^{[0]}(\xi)=-\frac{1}{2}\xi^2 \sigma^2~, \qquad A^{[1]}(\xi,D_\xi)=i\xi F(D_\xi)~. \nn
\eea
From Theorem~\ref{theorem1}, one needs to solve the recursion for $(j\geq 1)$
\bea
\rho_t^{[j]}(\psi(\xi))=\int_0^t e^{-\frac{1}{2}(t-s)\xi^2\sigma^2}
\Bigl\{i\xi F(D_\xi)\rho_s^{[j-1]}(\psi(\xi))ds+\Bigl(h_1 D_\xi+h_2\Bigr)\rho_s^{[j-1]}(\psi(\xi))dY_s\Bigr\}~\nn\\
\label{benes-asymptotic}
\eea
starting from the zeroth order solution:
\bea
\rho_t^{[0]}(\psi(\xi))=\exp\Bigl(-\frac{1}{2}t\xi^2\sigma^2\Bigr)~. \nn
\eea

\subsubsection{Polynomial-function approximation}
Here, we discuss how to obtain the polynomial approximation $F(x)$ for 
the nonlinear drift of the signal:
\be
f(x)=a\sigma \tanh(a x/\sigma)~. \nn
\ee 
Due to the normalization by $\sigma$, we can roughly expect
\be
\left|\frac{X_t}{\sigma}\right|\lesssim 1
\label{range}
\ee
for $t\in[0,1]$. This implies that Taylor expansion around $x=0$ is a
natural candidate for $F(x)$ when $|a|\lesssim 1$.
When $|a|\gtrsim 1$, the two plateaus of $f(x)$ start to play an important role in the range~(\ref{range}).
Unfortunately, however, Taylor expansion does not reproduce the plateaus but 
strongly diverging behavior within the range (\ref{range}) instead, which destabilizes the numerical result.
Thus, we take $[-5\sigma, 5\sigma]$ range with a step size of $0.2\sigma$, and carry out 
least-square method (LSM) to fit a $11$-dimensional odd function for $F(x)$.
We also adopt the weight function $g(x)$ defined as
\bea
g(x)=\exp\Bigl(-w\frac{x^2}{2\sigma^2}\Bigr) \nn
\eea
with various factor $w$. Here, $w=0$ corresponds to a pure LSM in the $5\sigma$ range and 
the polynomial function well recovers the two plateaus of $f(x)$  in a wide range, while it
has a relatively poor fit around the origin. On the other hand, higher $w$ gives finer fit and hence finer 
description of the density near $x=0$.  In this case, however, if one continues to increase $w$ it 
starts to destabilize the numerical result as in the case for Taylor expansion.
Thus, we need to take a balance of this trade-off, especially when $|a|>1$.

\subsubsection{Numerical Results}
In the following numerical examples, we take $t=1$ as the maturity and use $1,000$ steps to create
the sample observation (and signal) path. We then integrate (\ref{benes-asymptotic}) with the 
same time step $dt=10^{-3}$ for a given path of $Y$. For differentiation, we use a standard 
finite difference method. Finally, a discrete Fourier transformation is used to obtain 
the density function.

\begin{figure}[htb!]
\begin{center}	
\includegraphics[width=105mm]{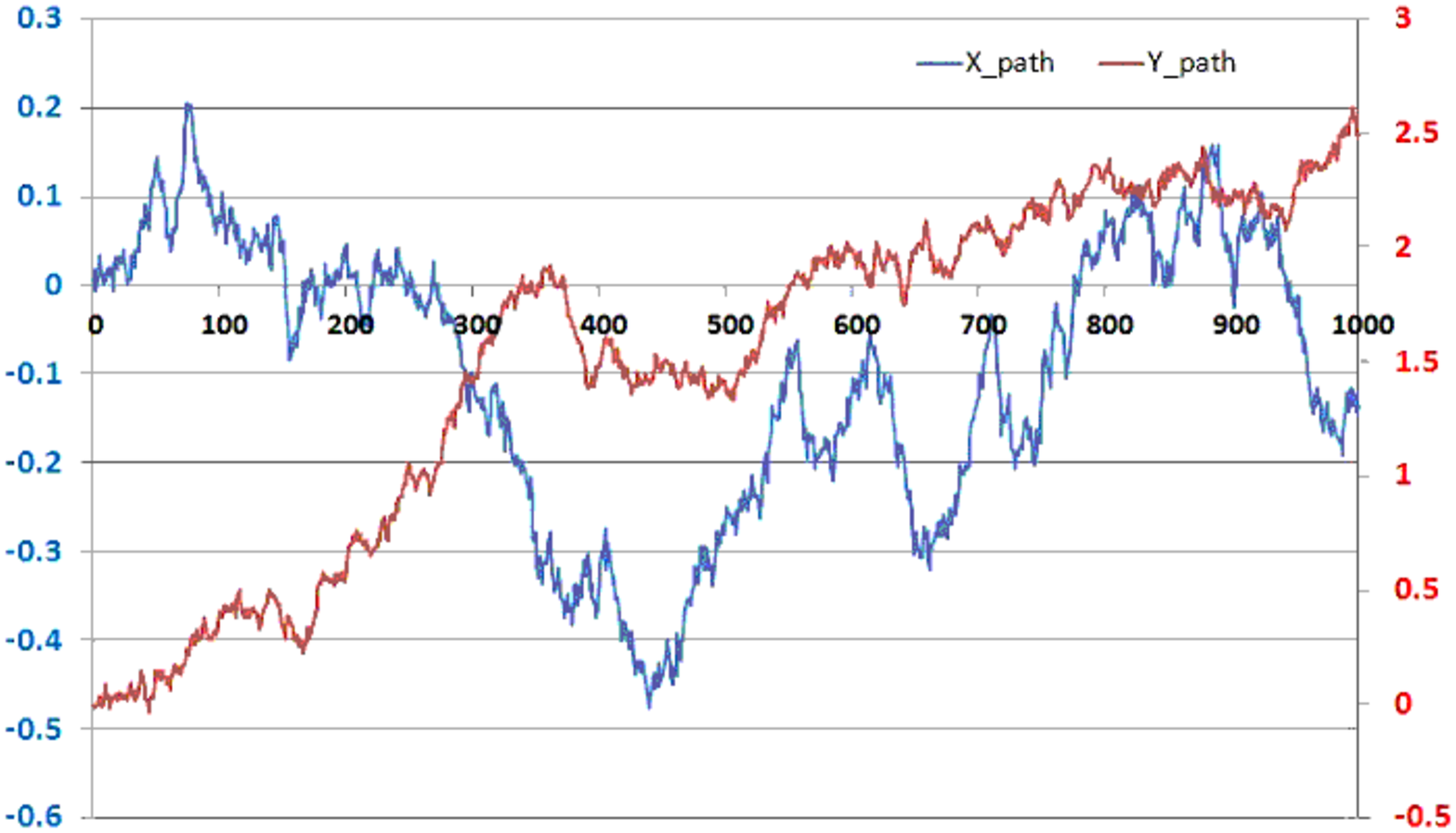} 
\includegraphics[width=105mm]{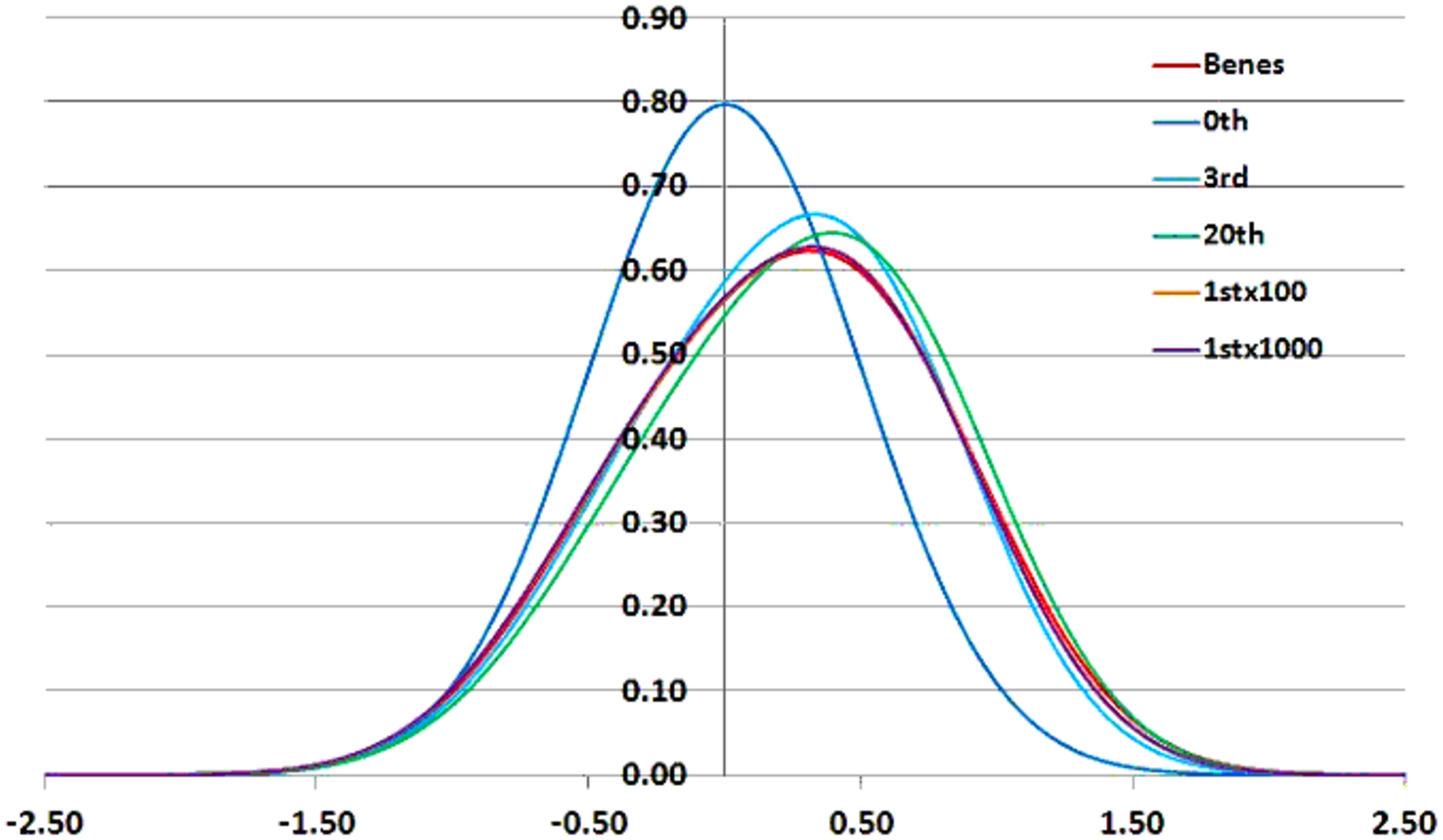} 
\includegraphics[width=105mm]{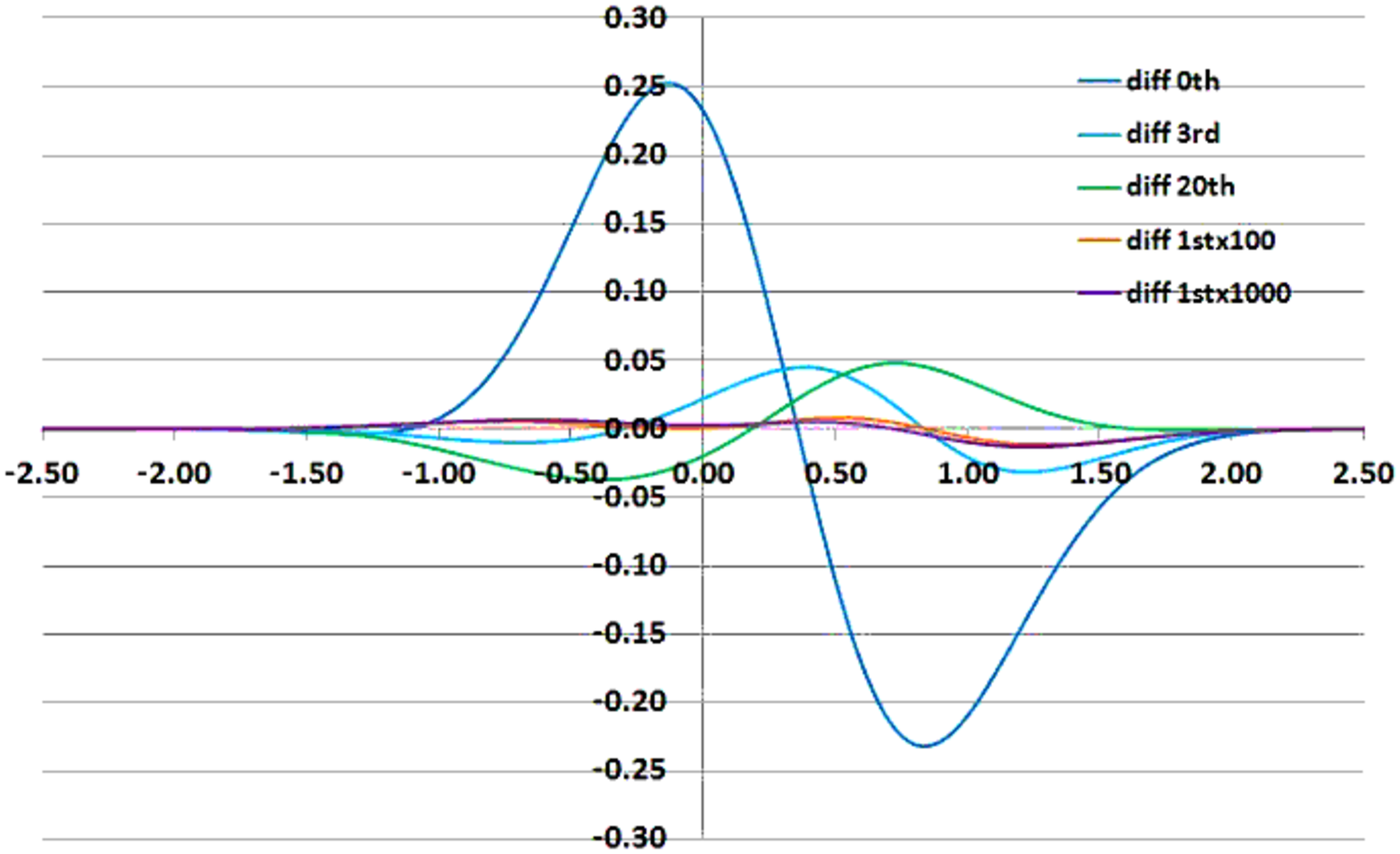}
\end{center}
\caption{$t=1, dt=10^{-3}, a=0.8, \sigma=0.5, h_1=0.8, h_2=0.5$ with polynomial function fitted with $w=2.0$. 
From top to bottom, the sample path,
exact and approximated density functions, and the difference of the approximated densities from the exact one.
In the middle graph, a red line labeled by "Benes" denotes the exact density function.}
\label{fig3}
\end{figure}

In the first numerical example given in Fig.~\ref{fig3}, we have used a set of parameters $\{a=0.8, \sigma=0.5, h_1=0.8, h_2=0.5\}$, and a sample path of $Y$ given in the top graph. We have used $w=2.0$
for getting coefficients of polynomial function $F(x)$.
The middle graph for the conditional density functions contains the exact one denoted by a red line labeled as "Benes",
estimated conditional densities from $(0th, 3rd, 20th)$-order asymptotic expansion without substepping method,
and those from 1st order expansion with substepping method of 
$(100, 1000)$ sub-periods~\footnote{First order expansion is good enough for 
short period since the infinitesimal generator contains no 2nd order term.}.
One can clearly see the benefit of substepping method explained in Sec.~\ref{substep}; Although there is no clear improvement from 3rd to 20th order approximation, the substepping method with small sub-period provides almost exact fit to the 
true density function.

\begin{figure}[htb!]
\begin{center}	
\includegraphics[width=105mm]{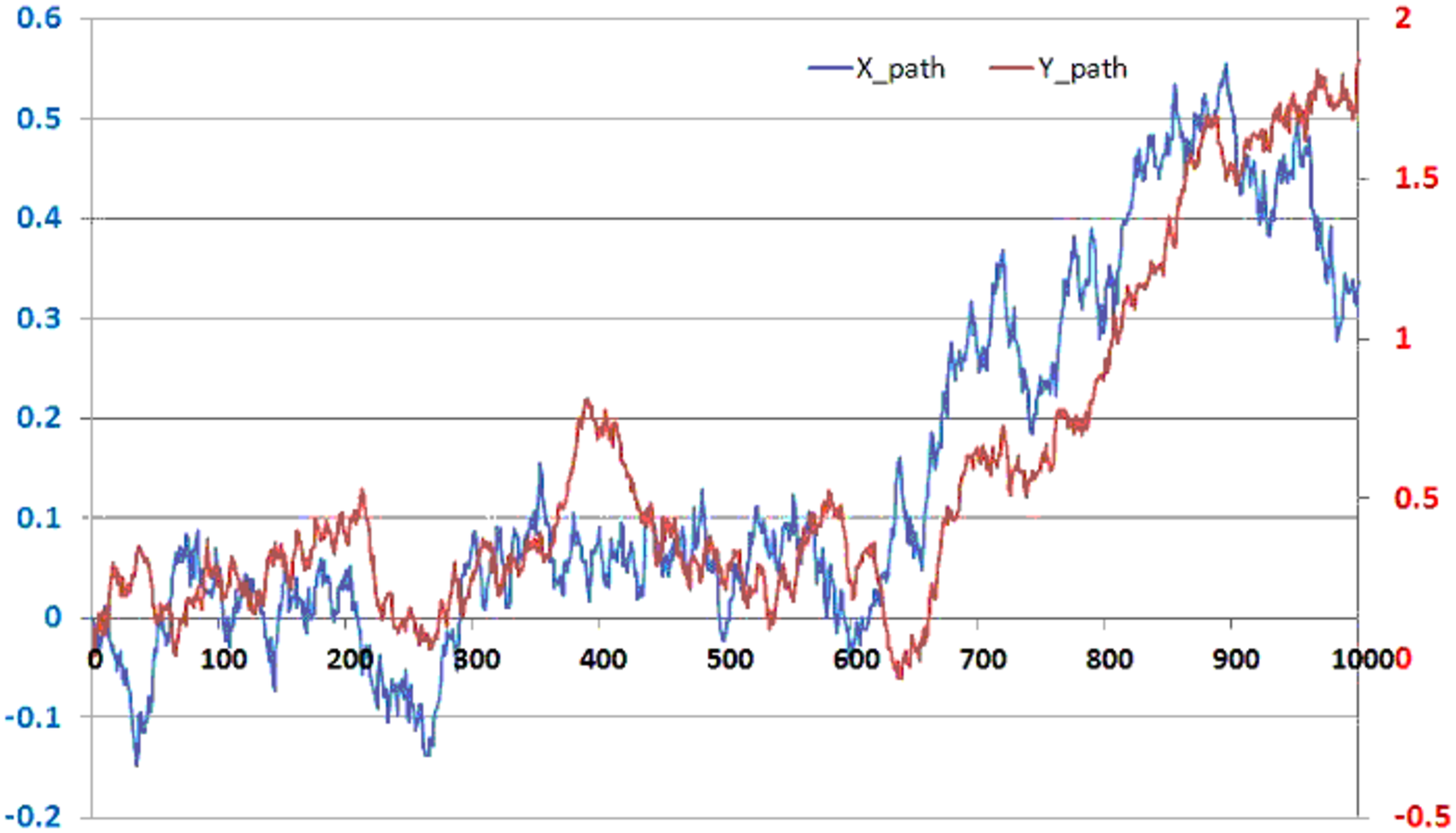} 
\includegraphics[width=105mm]{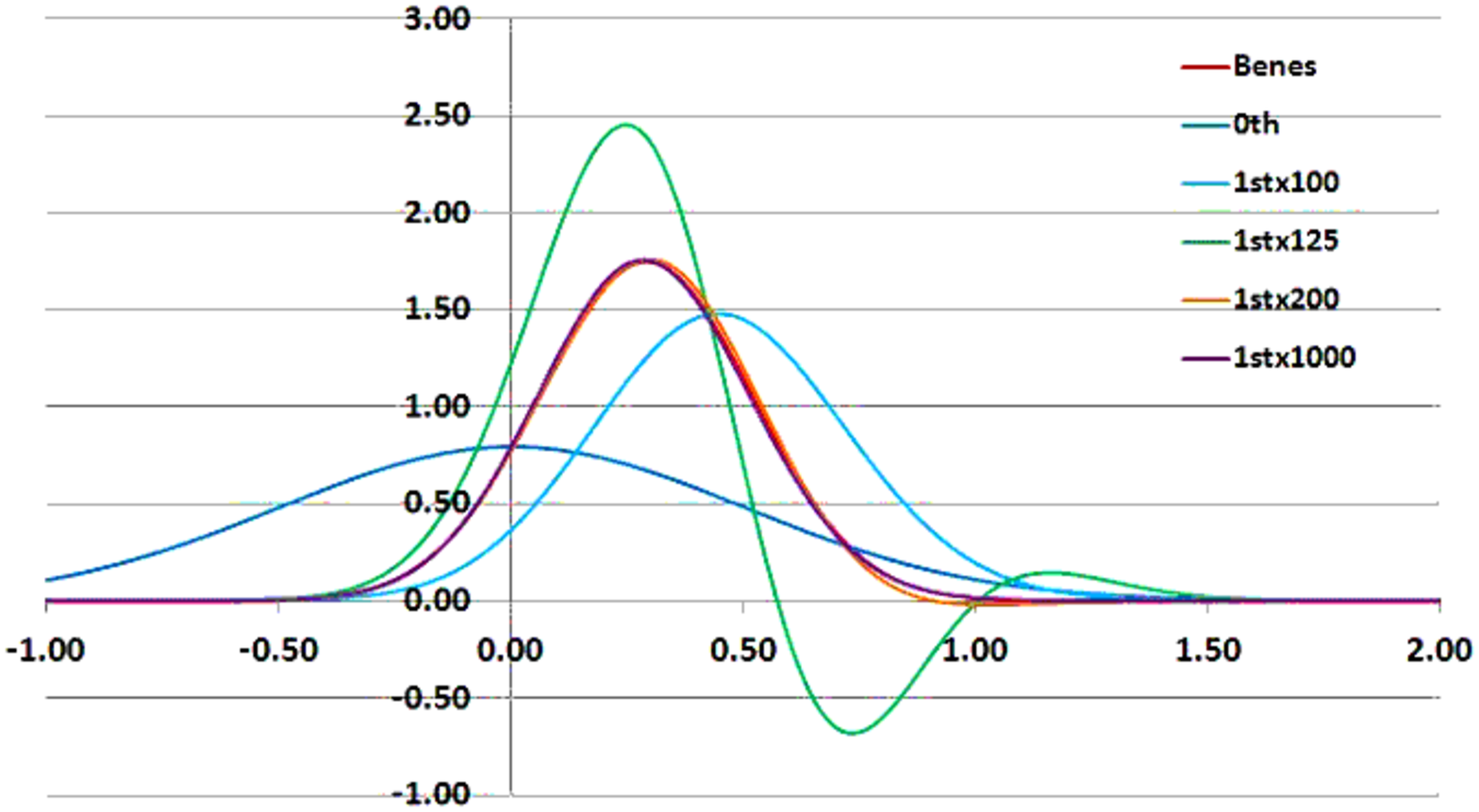} 
\includegraphics[width=105mm]{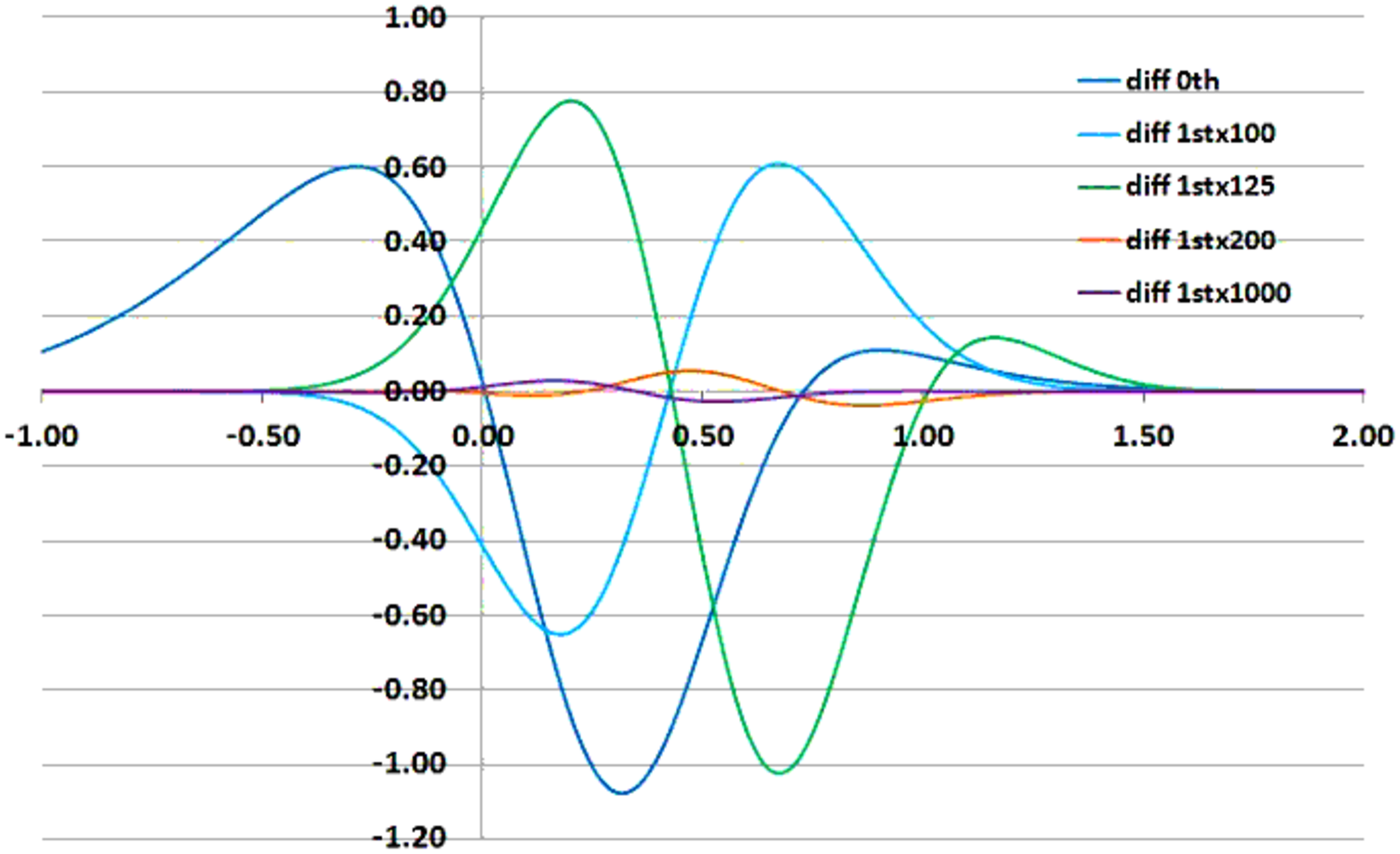}
\end{center}
\caption{$t=1, dt=10^{-3}, a=0.5, \sigma=0.5, h_1=10.0, h_2=0.5$ with polynomial function fitted with $w=2.0$. 
From top to bottom, the sample path,
exact and approximated density functions, and the difference of the approximated densities from the exact one.
In the middle graph, a red line labeled by "Benes" denotes the exact density function.}
\label{fig4}
\end{figure}

In Fig.~\ref{fig4}, we have used $h_1=10.0$, which is an example of  {\it small-noise} observation.
Since $a=0.5$ is relatively small, polynomial approximation for $f(x)$ is quite accurate ($w=2.0$).
The calculation has been performed with substepping method of (100, 125, 200, 1,000) sub-periods.
Fine substepping gives almost exact density even in this case. In particular, 
the significant reduction of the variance of the conditional density due to high quality information
provided by the observation process is well reproduced by the repeated application of asymptotic expansion.
In this example, approximation without substepping is too unstable and gives only meaningless 
numerical results.
As suggested in Sec.~\ref{substep}, the size of perturbation terms itself 
does not seem to be a relevant problem for asymptotic expansion {\it as long as} we have accurate enough polynomial function 
approximation and the substepping method.

\begin{figure}[htb!]
\begin{center}	
\includegraphics[width=105mm]{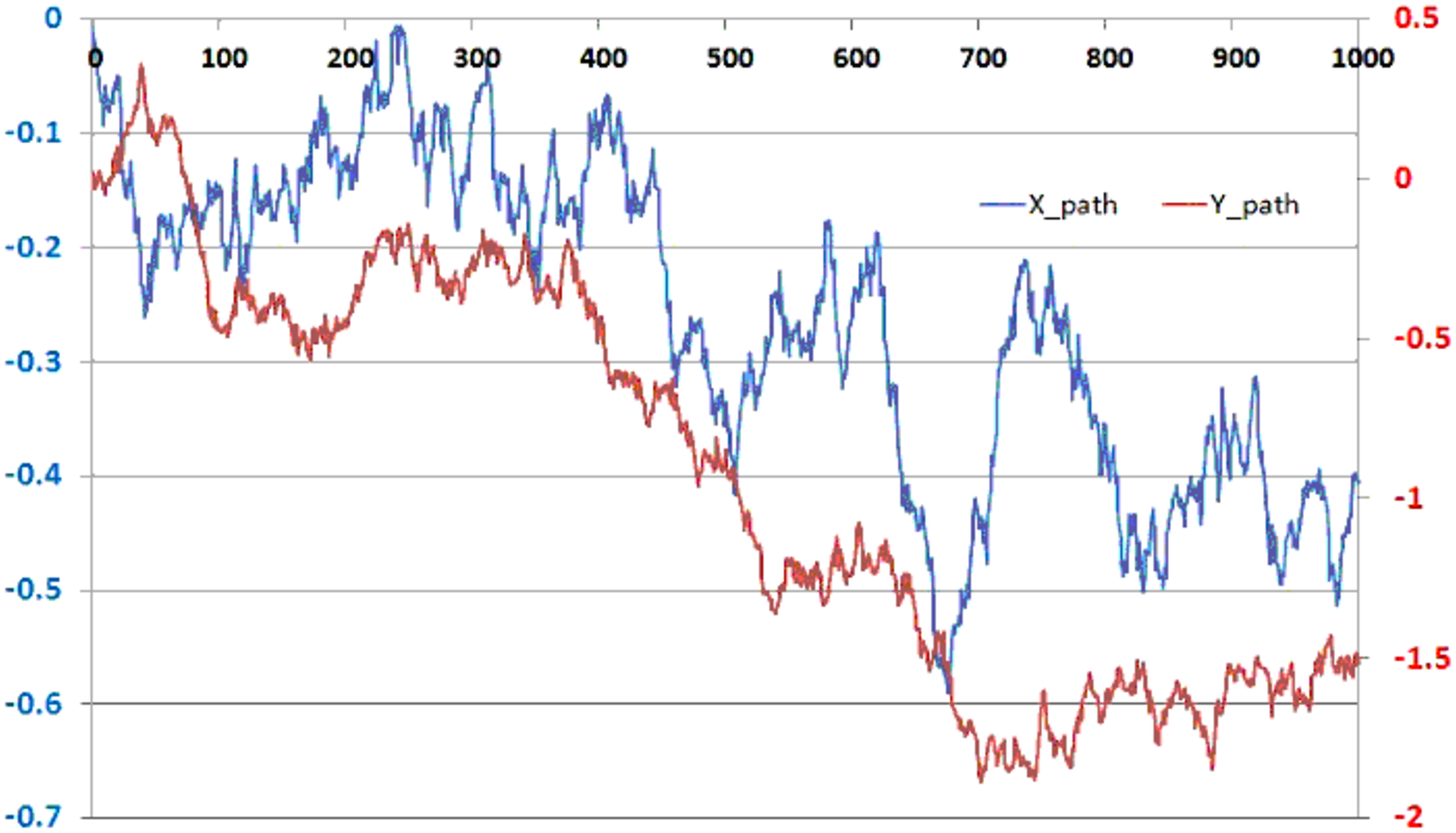} 
\includegraphics[width=105mm]{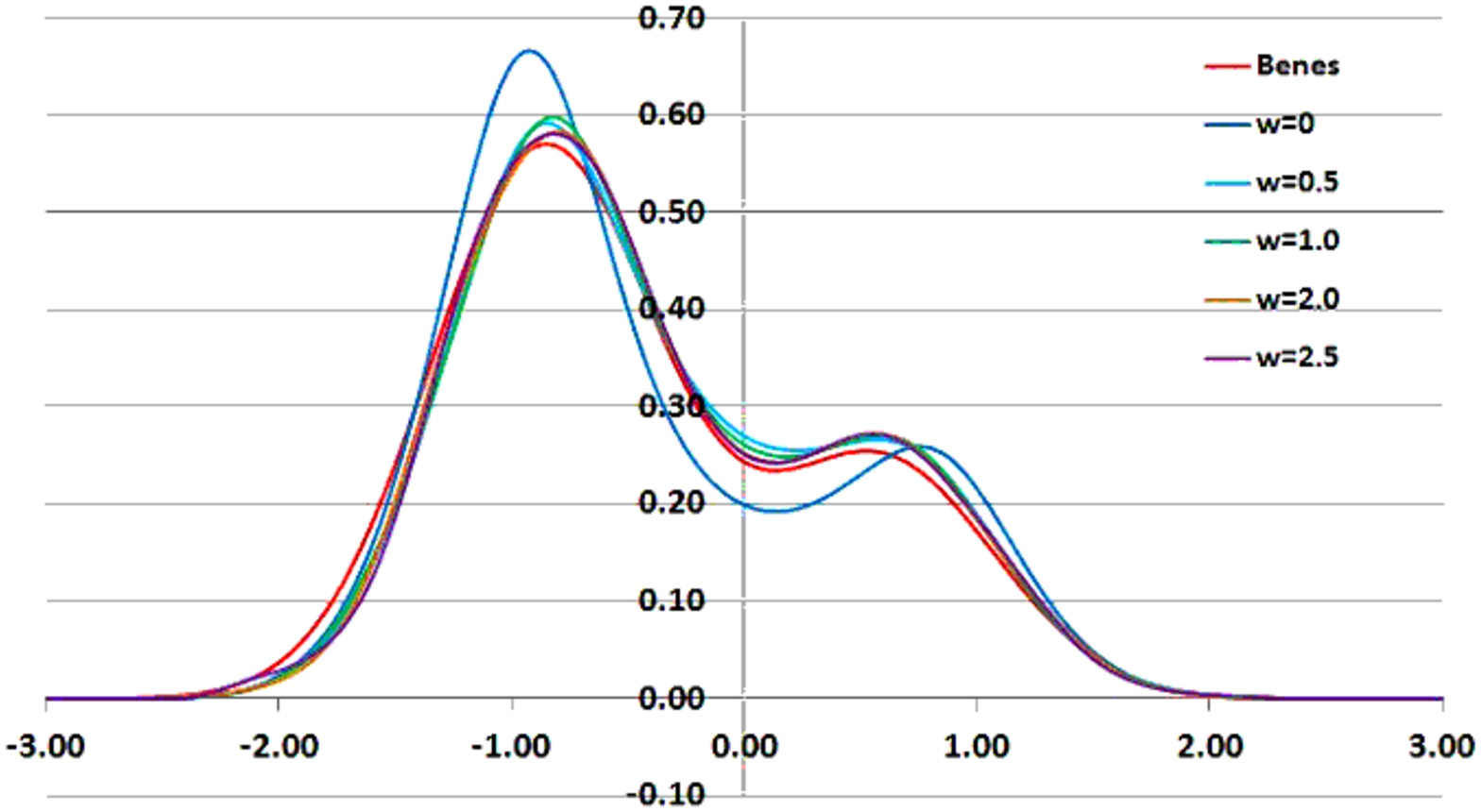} 
\includegraphics[width=105mm]{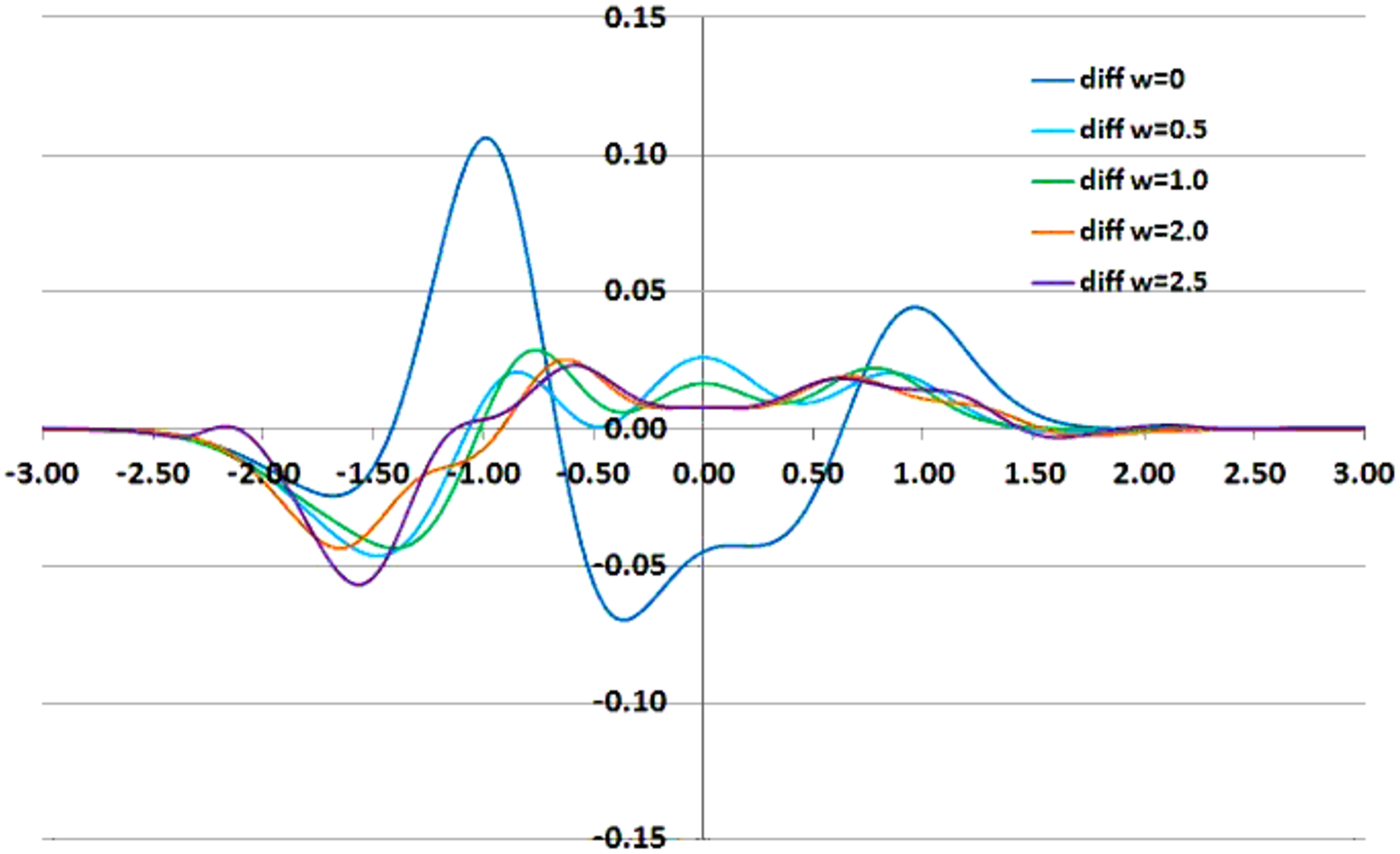}
\end{center}
\caption{$t=1, dt=10^{-3}, a=1.5, \sigma=0.5, h_1=0.7, h_2=0.5$ with $1,000$ substeps with 1st order asymptotic expansion. 
From top to bottom, the sample path,
exact and approximated density functions, and the difference of the approximated densities from the exact one.
In the middle graph, a red line labeled by "Benes" denotes the exact density function.}
\label{fig5}
\end{figure}

\begin{figure}[htb!]
\begin{center}	
\includegraphics[width=105mm]{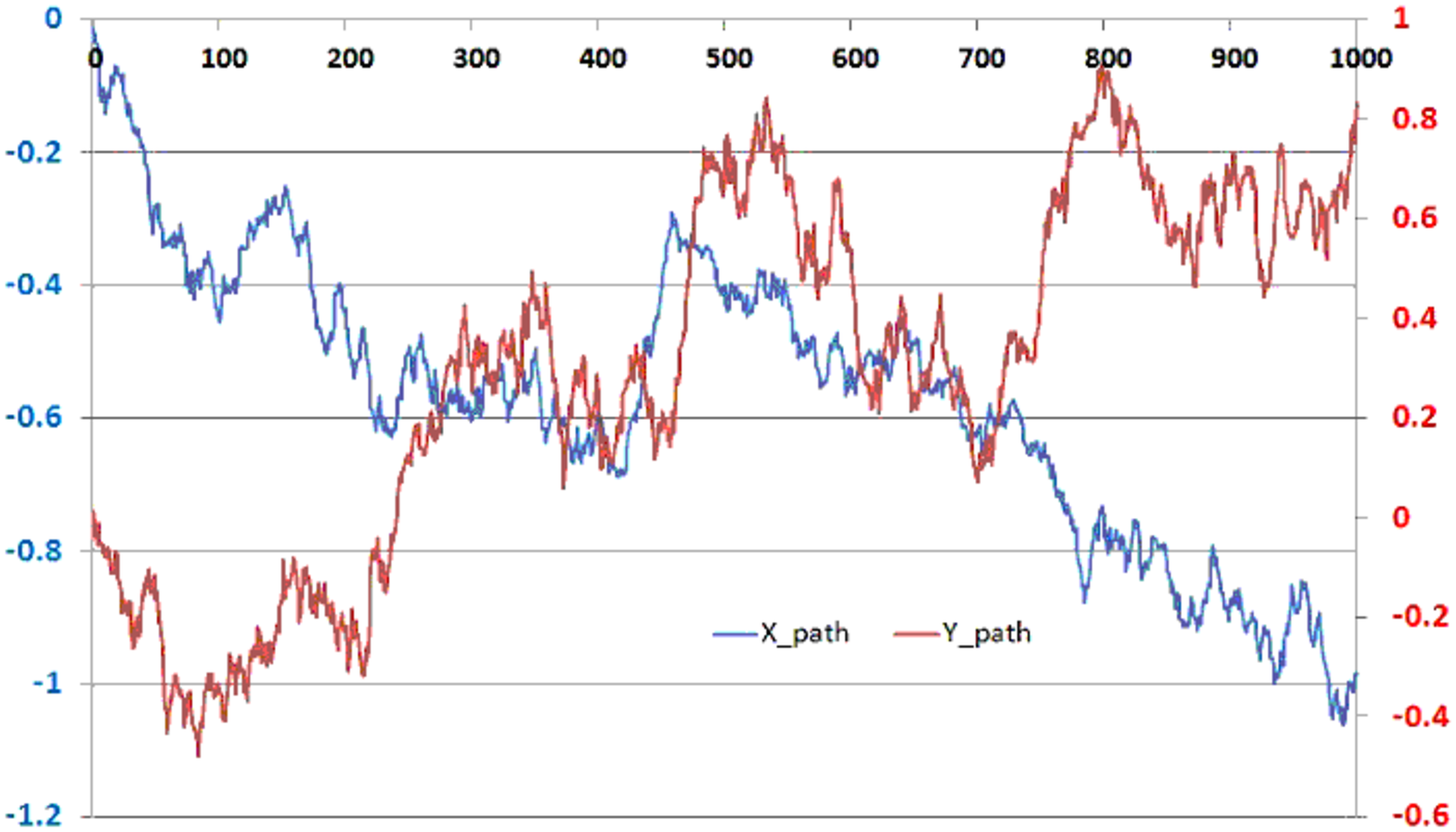} 
\includegraphics[width=105mm]{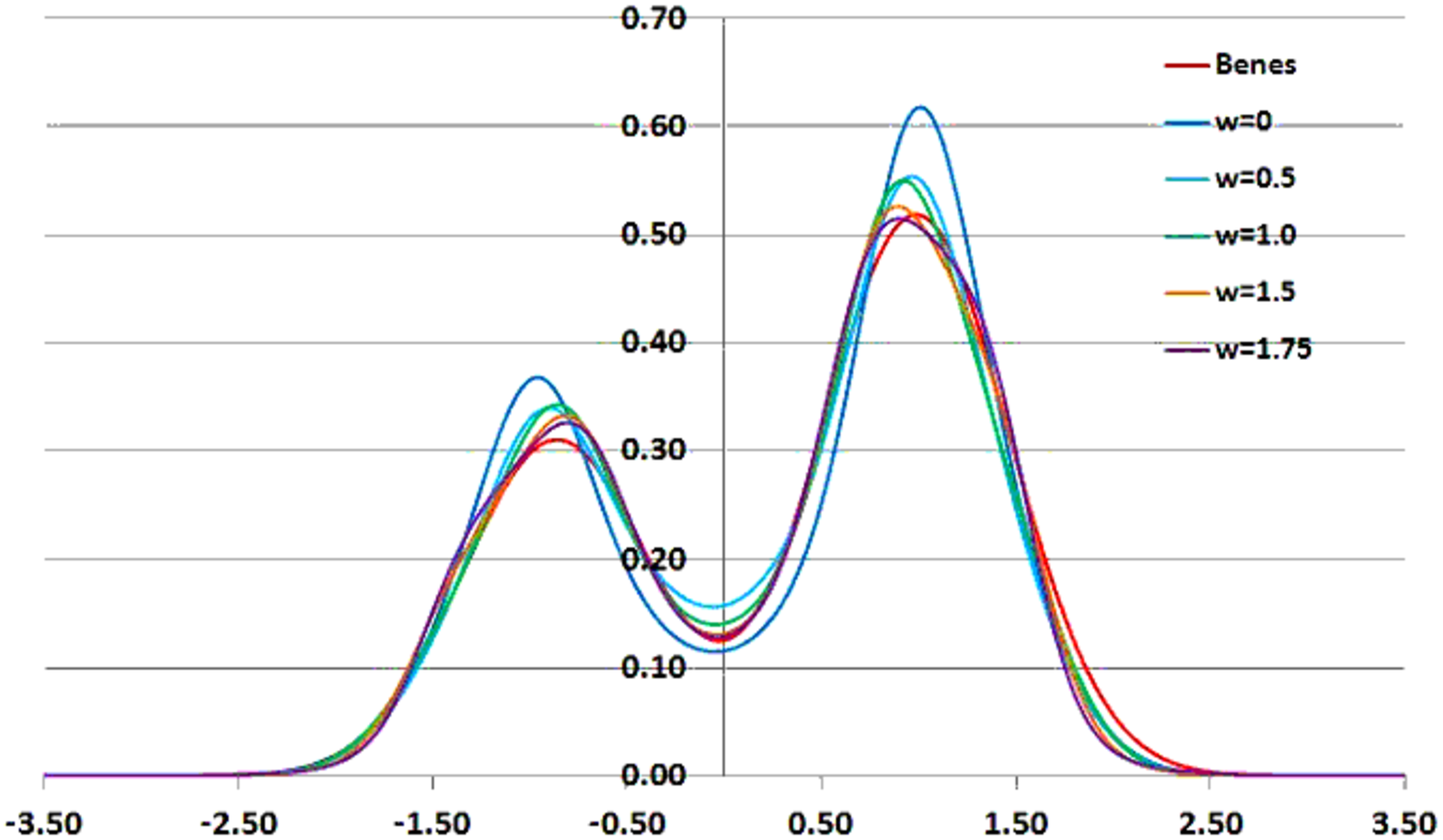} 
\includegraphics[width=105mm]{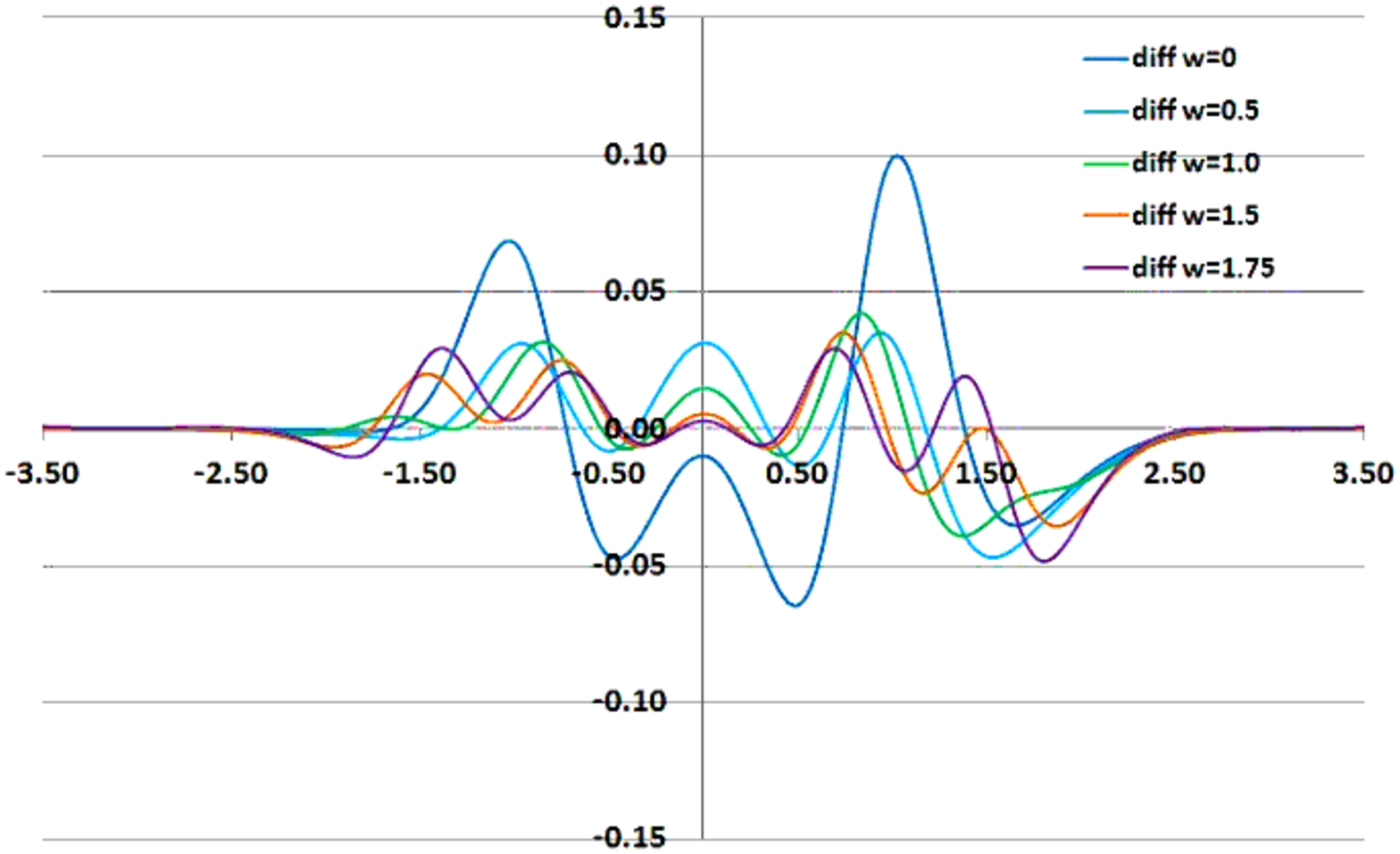}
\end{center}
\caption{$t=1, dt=10^{-3}, a=2.0, \sigma=0.5, h_1=1.0, h_2=0.5$ with $1,000$ substeps with 1st order asymptotic expansion. 
From top to bottom, the sample path,
exact and approximated density functions, and the difference of the approximated densities from the exact one.
In the middle graph, a red line labeled by "Benes" denotes the exact density function.}
\label{fig6}
\end{figure}

When one increases $a$, $f(x)$ becomes like a step function and makes it difficult to achieve accurate polynomial
approximation for the relevant range. Here, the choice of LSM weight $w$ starts to affect the estimated density.
In Fig.~\ref{fig5}, we have studied the case of $a=1.5$ with several choice of $w$.
Here, all the calculations were done with substepping method of $1000$ sub-periods.
The estimated density is stable for $w=0.5\sim 2.5$, but becomes unstable 
for higher $w$. Note that, the impacts of LSM weight are highly dependent on the order of the polynomial function.
As is easily guessed, the change of estimated density is more significant when lower order polynomials are used.
In Fig.~\ref{fig6}, we have done similar analysis for an example of $a=2.0$, which reveals more clearly 
separated two peaks of the filtered density. 
\\
\\
{\bf Remark :}\\
In the above  examples, we have used time-independent function for $F(x)$.
However, when there exists a significant drift for the signal process, such as large $|a|$
in the above example, making the polynomial function time dependent is quite likely to improve the approximation.
If we have the information about the evolution of the conditional mean, we can change the center of 
polynomial-function approximation to replicate the original nonlinear function more accurately in the relevant region.
Initial guess can be obtained by extended Kalman-Bucy filter or by time independent $F(x)$ in the current method, for 
example.

\section{Concluding Remarks}
In the paper, we have developed an asymptotic expansion technique in momentum space.  
Fourier transformation combined with polynomial-function approximation 
gives a closed recursive system of ODEs as an asymptotic expansion 
for the unnormalized conditional characteristic function.
Thanks to the simplicity of the ODE system, higher order calculation
can be performed easily. It also allows an efficient implementation of substepping method of 
asymptotic expansion. As long as polynomial approximation of the nonlinear terms is
accurate, the size of nonlinear terms ceases to be a big obstacle for obtaining an accurate estimation.
Applications to more realistic multi-dimensional filtering problems as well as 
other (financial) problems, such as option pricing (with some unobservable parameters),
are left for the future research. 
\\

Let us make a brief comment on the remaining problems and possible future directions of research to 
address these issues.
As one can see, the method still suffers from the curse of dimensionality. 
However, encouragingly, there exist a large number of works to ameliorate the higher dimensional integration problem.
See, for example, Griebel and Holtz (2010), Reisinger and Wittum (2007), Reisinger and Wissmann (2012) and Schr\"oder et al. (2012).  Especially, in Reisinger and Wissmann, the authors 
make use of the low effective dimensions of financial problems arising from a high correlation 
in the market. Although they have worked in restrictive model 
assumptions, they succeeded to approximate a high dimensional PDE by a series of low dimensional PDE.
Applications and improvement of these techniques by combining the asymptotic expansion 
developed in this paper look quite interesting.

Since we can only use finite order polynomials in practice, the quality of the estimation
highly depends on the accuracy of polynomial approximation.
When one has nonlinear terms difficult to fit by polynomials,
the idea of {\it change-of-variable} developed in Takahashi and Toda (2012) may be proved to be useful.
Suppose, one defines a new process $\wt{X}$ by using some function $\Psi(\cdot)$  as
\be
\wt{X}_t=\Psi(X_t)~. \nn
\ee
If $\wt{X}$ has drift and diffusion terms that are easier to be approximated by polynomial functions,
one can get more accurate estimation of
\bea
\wt{\rho}_t(\xi)=\mbb{E}\Bigl[e^{i\xi^\top \wt{X}_t}\Bigr] \nn
\eea
and hence also its density
\bea
\wt{\phi}_t(z)=\frac{1}{(2\pi)^n}\int_{\mbb{R}^n}e^{-i\xi^\top z}\wt{\rho_t}(\xi)d^n\xi~. \nn
\eea
Then, one can recover the density of the original $X_t$ by
\bea
\phi_t(z)=\wt{\phi}\Bigl(\Psi(z)\Bigr)\Bigl|J(z)\Bigr| \nn
\eea
where $|J|$ denotes the determinant of a Jacobian matrix with the elements of $(\part \Psi_i(z)/\part z_j)$.
Thus some of the errors can be absorbed if there exists an appropriate choice of $\Psi$.

Another possible solution is to use Fourier series expansion directly for nonlinear functions in the 
infinitesimal generator and observation process~\footnote{Note that, one have to resort to discrete Fourier transformation for numerical implementation anyway.}. Although it effectively increases the order of integration and hence slows down 
the calculation, some functions, such as step function, are known to allow accurate approximation by relatively 
a small number of terms. Suppose for example, some function $g$ has a Fourier expansion as
\bea
g(x)=\sum_n \wt{g}_n e^{-i\xi_n^\top x}~. \nn
\eea
where $\{\xi_n\}$ is a series of discretized momentum, and $\{g_n\}$ is a set of corresponding coefficients.
Then, one has
\bea
\mbb{E}[g(x)e^{i\xi_m^\top x}]=\sum_n \wt{g}_n \mbb{E}[e^{i(\xi_m-\xi_n)x}]~. \nn
\eea
Since all the nonlinear functions are included in the perturbation terms, one can write 
the dynamics of $\rho_t^{[j]}(\psi(\xi))$ in terms of $\{\rho_t^{[k]}(\psi(\xi_n)\}_n$ with $k=\{j-1, j-2\}$.
As a result, it is still linear for the highest expansion order and can be treated similarly as in Theorem~\ref{theorem1}.
This technique may be crucial for financial applications where stiff payoff functions are common.
Detailed studies of these points are also among our research topics in the future.

\section*{Acknowledgment}
The author thanks to Akihiko Takahashi of University of Tokyo for informative discussions that help
to understand the connections to the asymptotic expansion in the position space.
This research is supported by CARF (Center for Advanced Research in Finance) and 
the global COE program ``The research and training center for new development in mathematics.''




\clearpage

\end{document}